\documentclass[]{aa} 

\usepackage[normalem]{ulem}

\usepackage{graphicx}
\usepackage{txfonts}
\usepackage{siunitx}
\usepackage{longtable} 
\usepackage{lscape}
\usepackage{rotating}
\usepackage[]{natbib}
\usepackage{todonotes}
\usepackage{amsmath}
\usepackage{makecell}
\usepackage{multirow}
\usepackage[percent]{overpic}
\usepackage{subcaption}
\usepackage[colorlinks=true, allcolors=blue]{hyperref}

\DeclareSIUnit\gauss{G}

\begin{document}
\renewcommand{\arraystretch}{1.0}
\renewcommand{\nat}{Nat}

\title{Mitigating stellar radial velocity jitter\\ using orthogonal activity indices and\\a time-aware neural network}

\author{
Jordi Blanco-Pozo\inst{1,2,3}\thanks{\email{blanco@ice.csic.es}}
\and Manuel Perger\inst{1,2}
\and Guillem Anglada-Escudé\inst{1,2}
\and Ignasi Ribas\inst{1,2}
\and David Baroch\inst{2}
\and Marina Lafarga\inst{4,5}  
\and Juan Carlos Morales\inst{1,2} 
\and \`Oscar Porqueras-León\inst{1,2}
\and Sophie Stucki\inst{1,2}
\and David Vallmanya Poch\inst{6,7}
}

\authorrunning{J. Blanco-Pozo et al.}

\institute{
\inst{1}Institut de Ci\`encies de l'Espai (ICE, CSIC), Campus UAB, Carrer de Can Magrans s/n, 08193 Bellaterra, Spain\\
\inst{2}Institut d’Estudis Espacials de Catalunya (IEEC), Edifici RDIT, Campus UPC, 08860 Castelldefels, Barcelona, Spain\\
\inst{3}Facultat de Física, Universitat de Barcelona (UB), Martí Franqu\`es 1, 08028, Barcelona, Spain\\
\inst{4}Department of Physics, University of Warwick, Gibbet Hill Road, Coventry CV4 7AL, UK\\
\inst{5}Centre for Exoplanets and Habitability, University of Warwick, Coventry, CV4 7AL, UK\\
\inst{6}Delft University of Technology, Department of Imaging Physics, Gebouw 22, Lorentzweg 1, 2628 CJ Delft, Netherlands \\
\inst{7}Data Science Center, Barcelona School of Economics, Ramon Trias-Fargas 25-27, 08005 Barcelona, Spain
}

\date{Received \today; accepted \today}

\abstract
{Despite recent advances in the precision of high-resolution spectrographs, the detection of Earth-like exoplanets is still limited by the effects of stellar activity, which introduce radial velocity variations at the metre-per-second level or larger.}
{We present a framework to disentangle stellar effects from planetary signals by exploiting high-order distortions of the cross-correlation function (CCF; a measure of the average spectral line profile), thus moving beyond the commonly applied Gaussian fit approximation.}
{We decomposed the CCF using a Gram-Schmidt orthogonal basis function, enabling the separation of pure line shifts from line-shape distortions. To model activity-induced contributions to the radial velocities, we have developed a time-aware convolutional attention network dubbed CANSTAR. This network was trained on synthetic line-shape distortion coefficients produced with the realistic stellar simulator \texttt{StarSim} to learn the temporal evolution of stellar activity features.}
{We validated our framework using HARPS and CARMENES observations of two active stars, $\epsilon$\,Eridani and TZ\,Arietis. The network effectively mitigates stellar activity, reducing the radial velocity RMS to 52.5\,\% and 62.4\,\% of the uncorrected variability, respectively. This correction enables a more precise determination of the orbital parameters of TZ\,Arietis\,b compared to a Gaussian process regression.}
{Our results demonstrate that neural networks that incorporate the temporal context can outperform state-of-the-art methods in complex activity regimes. Future improvements on \texttt{StarSim} that will allow us to train CANSTAR on 3D magnetohydrodynamic spectra and more complex instrumental modelling are expected to bridge the performance gap between synthetic and real data, offering a robust pathway towards detecting Earth-mass planets around Sun-like stars.}

\maketitle 

\newcommand{\symb}[2]{$\mathrm{#1}_{\mathrm{#2}}$}
\newcommand{\symbmath}[2]{\mathrm{#1}_{\mathrm{#2}}}

\section{Introduction} \label{sec:intro}
Since the discovery of 51\,Pegasi\,b in 1995 \citep{mayor1995jupiter}, the radial velocity (RV) method has allowed for the detection of over a thousand exoplanets, moving over time from large Jupiter-like planets to smaller rocky worlds orbiting in the temperate zone of their host stars \citep[e.g.][]{hatzes2000evidence,mcarthur2004detection,anglada2016terrestrial}. This progress has been largely due to technical improvements on spectrographs, which have gradually pushed detection limits from signals in the tens of metres per second down to the m\,s$^{-1}$ domain, with instruments such as HARPS \citep{2003Msngr.114...20M}, HARPS-N \citep{cosentino2012harps}, CARMENES \citep{quirrenbach2016carmenes}, and SPIRou \citep{donati2018spirou}. The newest spectrographs, for example, ESPRESSO \citep{espresso2010}, with an instrumental precision reaching down to around 10\,cm\,s$^{-1}$, represent a big leap towards the detection of even smaller Keplerian signals induced by Earth-like rocky planets \citep{figueira2025comprehensive}.

Despite these instrumental advancements, stellar magnetic activity remains a fundamental barrier to detecting low-amplitude RV signals. Surface inhomogeneities such as spots, faculae, and granulation can induce RV variations greater than 1\,m\,s$^{-1}$, even in relatively quiet stars \citep[e.g.][]{alphaCen2012,perger2017hades}. These stellar activity effects distort spectral line shapes, potentially mimicking or obscuring planetary signals. The two main RV extraction methods are affected by these wavelength-dependent distortions, and the community has developed different strategies for each technique to disentangle planetary signals from instrumental or stellar activity variability.

In the cross-correlation function (CCF) method, the observed spectrum is cross-correlated with a weighted mask tailored to the stellar spectral type \citep{baranne1996elodie}. The resulting CCF represents an average stellar absorption line. For most stars, the CCF is well described by a Gaussian profile. The centroid yields the RV, while the contrast (CON) and the full width at half maximum (FWHM $= 2\sqrt{\ln 2}\,\sigma$, where $\sigma^2$ is the variance of the Gaussian) are line-shape activity indicators. The bisector inverse slope (BIS), defined as the velocity difference between the upper (60–90\,\%) and lower (10–40\,\%) parts of the CCF, is used as a measurement of line asymmetry \citep{queloz2001no}. Several alternative approaches have been developed to extract activity information from the CCF, including bi-Gaussian fitting with asymmetric widths \citep{figueira2013line}, Fourier decomposition \citep{zhao2020fiesta}, principal component analysis (PCA) of the auto-correlation function \citep{collier2021separating}, and PCA on shape-driven CCFs orthogonalised with respect to the first derivative of a template \citep{klein2024investigating}.

In spectral-level methods, RVs are computed directly from the observed spectra using high signal-to-noise templates. The optimal wavelength shift that aligns an observation with the template is typically obtained through one of two main formalisms: (i) a numerical approach that performs a least-squares minimisation by exploring the velocity parameter space or (ii) an analytical approach based on the \citet{bouchy2001fundamental} formalism, where the shift is inferred from a first-order Taylor expansion of the spectrum. The latter requires calculating the derivative of the flux with respect to wavelength from a nearly noise-free template. Either of these formalisms can be applied over different wavelength ranges, thus broadly dividing their practical application into two techniques. (1) Global template matching applies the shift measurement to the entire spectrum at once \citep{anglada2012harps, zechmeister2018spectrum,silva2022novel}. Line-width variations are captured in the differential line width activity indicator \citep{zechmeister2009generalised}, and the difference in RV variability as a function of wavelength is measured with the chromatic index \citep{zechmeister2018spectrum, baroch2020carmenes}. (2) The line-by-line RV method allows us to measure the RV signal from each individual line with respect to the template, rather than only producing a single global RV value \citep{dumusque2018measuring, artigau2022line, lafarga2023carmenes}. This allows for detailed studies of differential line responses to activity, making it possible to select subsets of lines based on their line-by-line RV sensitivity to parameters such as line depth \citep{cretignier2022stellar}, formation temperature \citep{al2022measuring}, or spot-to-photosphere temperature contrast \citep{larue2025chromaticity}. More data-driven approaches, such as PCA-based line selection, have also been explored \citep{cretignier2023yarara}.

Several techniques have been developed to correct for stellar activity using either CCF-based or spectral-level diagnostics. Gaussian process (GP) regression has become a widely adopted method for modelling the quasi-periodic variability induced by stellar magnetic activity in RV time series \citep{haywood2014planets,ambikasaran2015fast,foreman2017fast, perger2021auto}. More recently, multi-dimensional GP frameworks have been introduced, combining RVs with ancillary indicators such as photometry, CCF diagnostics, or chromospheric line indicators \citep[e.g.][]{barragan2022pyaneti, delisle2022efficient}. Neural network (NN)-based methods have also gathered significant interest. Convolutional neural networks (CNNs) have been applied to solar CCFs \citep{debeurs2022identifying} and to other input data, such as the spectral-shell representations \citep{cretignier2022stellar,zhao2024improving}, though without explicit temporal modelling. \citet{perger2023machine} extended CNNs to time series modelling of CCF indicators using physically motivated training data generated with the \texttt{StarSim} code \citep{herrero2016modelling}. Architectures such as autoencoders have also been proposed, for example, to distinguish real from apparent spectral-line shifts in simulated data \citep{liang2023aestra}.

In this work, we propose an alternative approach based on separating line shifts from line shape distortions with an orthogonal basis function whose derived activity indicators are used to train a time-aware NN to mitigate stellar activity in the RV data. We apply the methodology to two test stars: $\epsilon$\,Eridani ($\epsilon$\,Eri) and TZ\,Arietis (TZ\,Ari). These represent two distinct cases with different CCF shapes, so we aim to prove the capability of our method to derive activity indicators on CCFs with Gaussian and non-Gaussian shapes. 

The paper is organised as follows. The test stars are presented in Sect.~\ref{sec:teststar}. In Sect.~\ref{sec:Theoretical_framework}, we introduce the method to extract line-shape-indicator coefficients from the CCF, and in Sect.~\ref{sec:Application_method} we apply this theoretical framework to the test cases. In Sect.~\ref{sec:convatt} we explain how we fed these indicators into a NN architecture for time series modelling (convolutional attention network) and compare its performance to a network agnostic to time information (fully connected network). We trained both networks on synthetic data produced with the {\tt StarSim} code. In Sect.~\ref{sec:Results}, we evaluate the best-performing network on the two test cases and study the improved sensitivity to planetary signals provided by the network's stellar activity mitigation. In Sect.~\ref{sec:discussion} we discuss the results, and in Sect.~\ref{sec:Conclusions} we summarise the main findings and implications of the study.

\section{Test cases $\epsilon$\,Eri and TZ\,Ari} \label{sec:teststar}

The decomposition framework of this study was validated on two test stars. These specific targets were chosen to showcase the framework's ability to capture both Gaussian-like and non-Gaussian-like CCF profiles.

\subsection{Target characteristics}

\subsubsection{$\epsilon$\,Eri}

$\epsilon$\,Eri is a young \citep[400--800 Myr;][]{mamajek2008improved} K2 dwarf located at a distance of 3.22\,pc, making it one of the closest known exoplanet hosts. With a rotation period of 11.2\,d \citep{frohlich2007differential} and $v\sin{i} = 2.4 \pm 0.5$\,km\,s$^{-1}$ \citep{valenti2005spectroscopic}, it serves as a representative example of a moderately active star of this age. The system hosts a prominent debris disk \citep{greaves1998dust} and a Jupiter-mass planet, $\epsilon$\,Eri\,b, with an orbital period of approximately seven years \citep{hatzes2000evidence, llop2021epserib}. \citet{giguere2016combined} estimated a stellar inclination of the rotation axis to be $69.5^{+5.6}_{-7.6}$\,deg by modelling spot modulation in photometric and RV data. We summarise the relevant stellar parameters in Table~\ref{tab:stellar_params}.

For this study, we used 205 publicly available spectra from the HARPS spectrograph, on the ESO 3.6\,m telescope at La Silla Observatory, Chile. HARPS covers the optical wavelength range from 0.38 to 0.69\,$\mu$m at a resolving power of $R \approx 120{,}000$. Our dataset covers a baseline of 88 days between October~5,~2019 (JD~2458762) and January~1,~2020 (JD~2458850). To mitigate short-term variations, we binned the observations nightly, resulting in 66 final data points.

\begin{table*}[]
\caption{Stellar parameters and adopted values for the \texttt{StarSim} simulations for $\epsilon$\,Eri and TZ\,Ari. \label{tab:stellar_params}}
\centering
\begin{tabular}{l|lll|lll}
\hline \hline
Parameter & Literature & Ref & \texttt{StarSim} value & Literature & Ref & \texttt{StarSim} value \\ \hline 
Name & $\epsilon$ Eri & & -- & TZ\,Ari (Gl\,83.1) & & -- \\
Spectral type & K2.0\,V & (1) & \makecell[l]{HARPS K5 mask} & M5.0\,V & (2) & \makecell[l]{CARMENES M5 mask} \\
$\alpha$ & 03:32:54.79 & (3) & -- & 02:00:14.16 & (3) & -- \\
$\delta$ & $-$09:27:29.41 & (3) & -- & +13:02:38.66 & (3) & -- \\
$\mu_{\alpha}$ (mas\,yr$^{-1}$) & $-$974.76$\pm$0.16 & (3) & -- & 1096.46 $\pm$0.07 & (3) & -- \\
$\mu_{\delta}$ (mas\,yr$^{-1}$) & 20.88$\pm$0.12 & (3) & -- &   $-$1771.53$\pm$0.06 & (3) & -- \\
$d$ (pc) & 3.2198$\pm$0.0014 & (3) & -- &       4.4696$\pm$0.0014 & (3) & -- \\
$G$ (mag) & 3.4658$\pm$0.0014 & (3) & -- & 10.6811$\pm$0.0008 & (3) & -- \\
$L$ (L$_\odot$) & 0.32$\pm$0.05 & (4) & -- & 0.00254$\pm$0.00002 & (5) & -- \\
age ($Myr$) & 400-800 & (6) & -- & >750 & (7) & -- \\
$M$ (M$_\odot$) & 0.82$\pm$0.05 & (4) & 0.815--0.825 & 0.140--0.160 & (5) & 0.150$\pm$0.010 \\
$R$ (R$_\odot$) & 0.74$\pm$0.01 & (4) & 0.73--0.75 & 0.164$\pm$0.005 & (5) & 0.163--0.165 \\
$T_{\rm eff}$ (K) & 5076$\pm$30 & (8) & 4900–5300 & 3154$\pm$54 & (9) & 2950--3350 \\
$P_{\rm rot}$ (days) & 11.2 & (10) & 11.1-11.3 & 1.96$\pm$0.07 & (11) & 1.9--2 \\
$i$ (deg) & 69.5$^{+5.6}_{-7.6}$ & (12) & 70--90 & <28 & (7) & 20--28 \\
$\log{g}$ (dex) & 4.30$\pm$0.08 & (13) & 4.5 & 5.01$\pm$0.06 & (9) & 5 \\
T$_{\rm con}^*$ (K) & 1080$\pm$670 & (14) & 200--1000 & 536$\pm$318 & (14) & 200--1000 \\

CB$_\odot^*$ & $\sim$0.3 & (15) & 0.1--0.5 & -- & -- & $-$0.2--0.2 \\
\hline
\end{tabular}
\label{tab:stellar_params}

\tablefoot{ 
(1) \citet{keenan1989perkins};  
(2) \citet{lepine2013spectroscopic};
(3) \citet{brown2021gaia};
(4) \citet{baines2011confirming};
(5) \citet{schweitzer2019carmenes};
(6) \citet{mamajek2008improved};
(7) \citet{quirrenbach2022carmenes}; 
(8) \citet{heiter2015gaia}; 
(9) \citet{passegger2019carmenes}; 
(10) \citet{frohlich2007differential};
(11) this work;
(12) \citet{giguere2016combined}; 
(13) \citet{gonzalez2010parent}; 
(14) \citet{herbst2021starspots}; 
(15) \citet{liebing2021convective}.\\
$^*$ T$_{\rm con}$ is the spot temperature difference with respect to the photosphere; CB$_\odot^*$ the convective blueshift in solar units.
}
\end{table*}

\subsubsection{TZ\,Ari}

TZ\,Ari (Gl\,83.1) is a nearby M5.0 dwarf located at 4.47\,pc from Earth. \citet{quirrenbach2022carmenes} reported a rotation period of 1.96\,d based on spectroscopic indicators and ground-based photometric data from multiple facilities. This is confirmed by our analysis of TESS photometry \citep{ricker2015transiting} from September to November 2023 (Appendix~\ref{sec:TESS}). The rotation period combined with the upper limit on the projected rotational velocity $v\sin{i} < 2$\,km\,s$^{-1}$ \citep{reiners2018carmenes} implies an upper limit on the stellar inclination of $28^{\circ}$. \citet{quirrenbach2022carmenes} also identified a Saturn-mass planet, TZ\,Ari\,b, with an orbital period of 772.05$^{+2.41}_{-1.84}$\,d, a minimum mass of $0.21 \pm 0.02\,M_{\mathrm{Jup}}$, and an orbital eccentricity of $e = 0.46^{+0.06}_{-0.07}$. 

For this work, we used 92 spectra (after discarding three clear outliers as identified on the CCF activity indicators time series) obtained with the high-resolution CARMENES spectrograph on the 3.5\,m telescope at the Calar Alto Observatory. We employed CARMENES visual channel, which covers the optical wavelength range from 0.52 to 0.96\,$\mu$m at a resolving power of $R \approx 94{,}600$. The observations span 92 nights between January~31,~2016 (JD~2457419), and November~29,~2019 (JD~2458817). The last observing season, starting on July~17,~2019 (JD~2458682), was conducted at a higher cadence to properly sample the stellar rotation period.

\subsection{Spectroscopic data} \label{ssec:specdat}

\begin{figure} 
    \centering
    \includegraphics[width=\columnwidth]{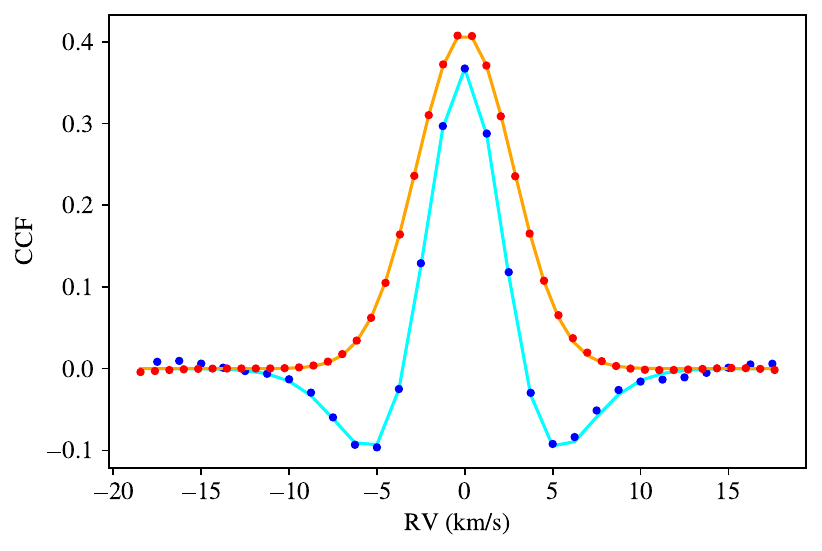}
    \caption{Time-averaged CCFs (templates) of $\epsilon$\,Eri (red) and TZ\,Ari  (blue) with the corresponding Gaussian model fit for $\epsilon$\,Eri  (orange) and the two-component Gaussian model fit for TZ\,Ari (cyan).}
    \label{fig:CCF_basis_function_fit_EpsEri_TZ_Ari}
\end{figure}

The data for each target star were analysed with the CCF method using the \texttt{raccoon} pipeline \citep{lafarga2020carmenes}. We employed masks tailored to the spectral type of each star (Table~\ref{tab:stellar_params}), adopting the detector pixel size as the velocity sampling step for the CCF computation. We computed the templates for each target star by time-averaging the flux-normalised CCFs. We normalised the flux by dividing each CCF by its baseline continuum and subtracting the result from unity, ensuring that the final profiles act as strictly positive, probability-like functions. The final templates are shown in Fig.~\ref{fig:CCF_basis_function_fit_EpsEri_TZ_Ari}. For the $\epsilon$\,Eri template (red), a Gaussian function provides an accurate description of the shape. The TZ\,Ari template (blue) shows a Gaussian shape in the core of the CCF, but displays two humps at either side, which is typical of CCFs of cool M dwarfs measured at visible wavelengths  \citep{lafarga2020carmenes}. General characteristics of the RV, CON, FWHM and BIS time series for both targets are presented in the table of Appendix~\ref{sec:appendix_table}.

\subsection{Basic data analysis}

To identify periodicities in the time series, we employ the Generalised Lomb–Scargle \citep[GLS, ][]{zechmeister2009generalised} periodogram. The significance of the detected peaks is assessed using false-alarm probability (FAP) levels derived via bootstrap resampling \citep{efron1985bootstrap}. We classify signals as tentative or significant if their power exceeds the 1\,\% or 0.1\,\% FAP levels, respectively. 

\begin{figure} 
    \centering
    \includegraphics[width=0.5\textwidth]{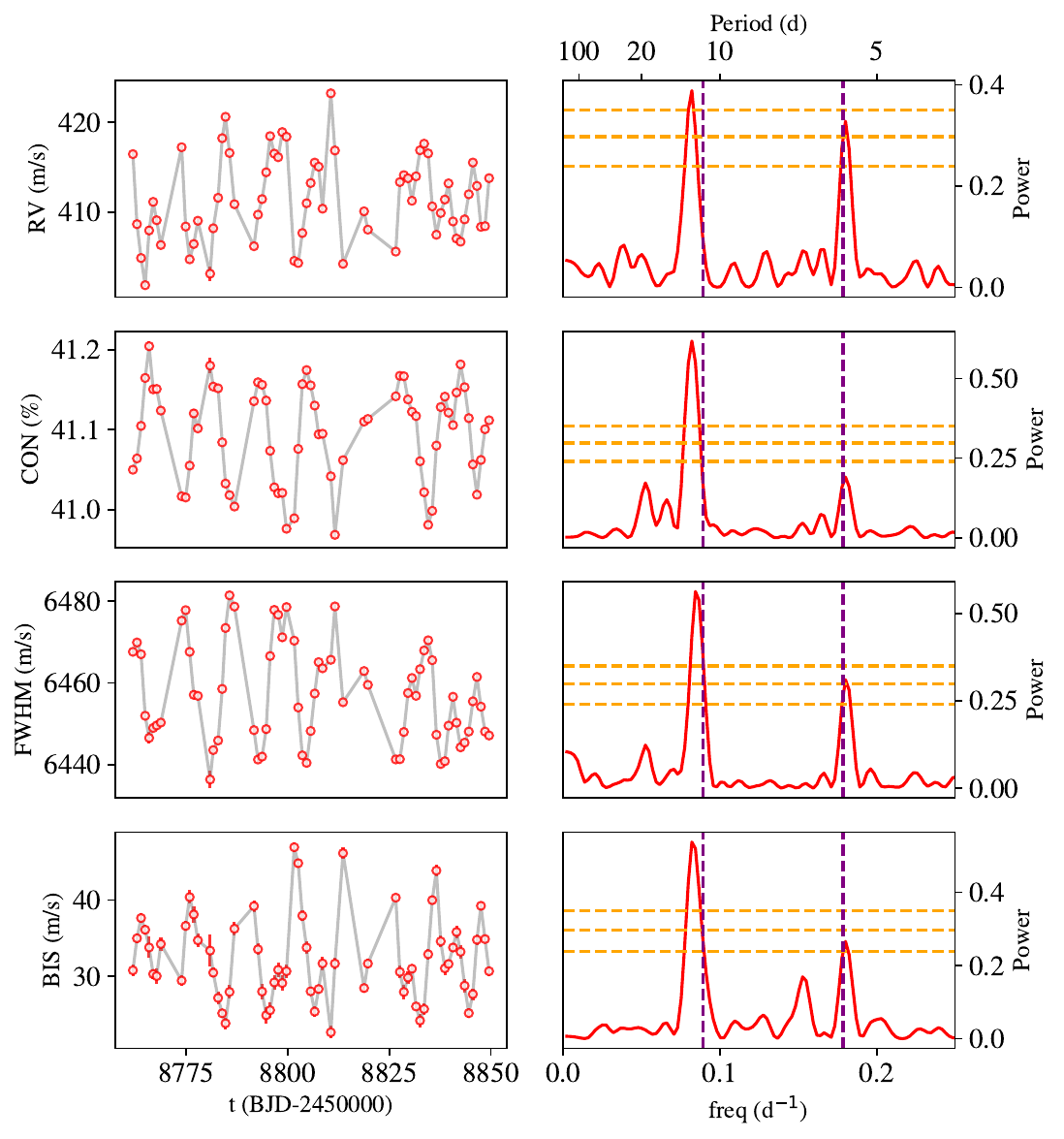}
    \caption{Time series (left) and GLS periodograms (right) of the RV and activity indicators derived from the CCFs for $\epsilon$\,Eri. We mark in purple the stellar rotation period (11.2\,d) and its first harmonic (5.6\,d). The 0.1, 1, and 10\,\% FAP levels are shown with dashed orange lines.}
    \label{fig:EpsEri_ClassicalActivity_TimeSeries_LS}
\end{figure}

\begin{figure} 
    \centering
    \includegraphics[width=0.5\textwidth]{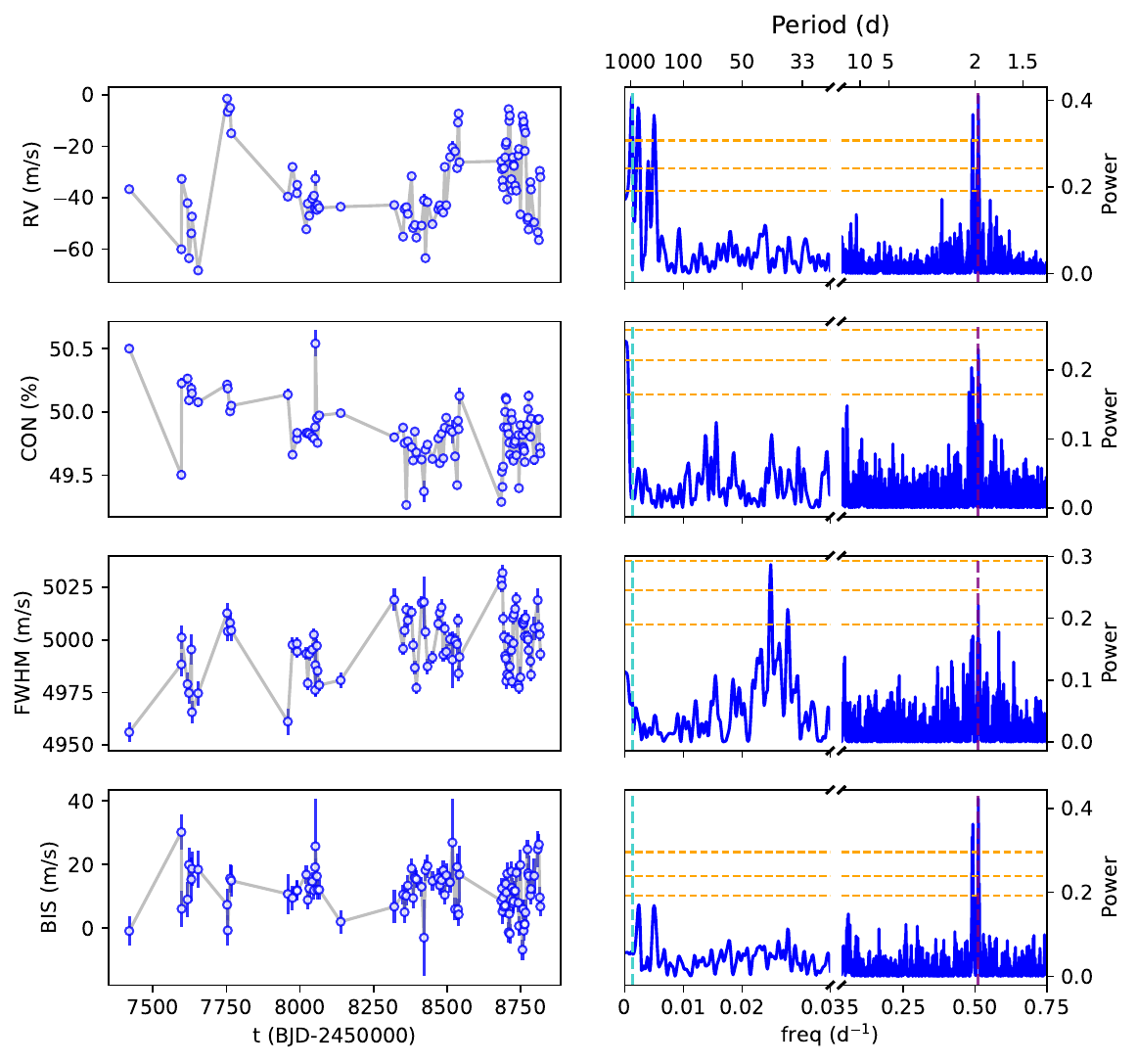}
    \caption{Time series (left) and GLS periodograms (right) of the RV and activity indicators derived from the CCF for TZ\,Ari. The stellar rotation period (1.96\,d) is marked in purple; the planetary orbital period from the literature \citep[722.05\,d;][]{quirrenbach2022carmenes} is marked in light blue; and the 0.1\,\%, 1\,\%, and 10\,\% FAP levels are shown with dashed orange lines. The frequency axis of the periodogram show two scales split at 0.035\,d$^{-1}$ to better visualise long-term variability. The RVs display low-frequency signals associated with the Keplerian motion of TZ\,Ari~b, while the CON and FWHM show lower-frequency stellar activity signals.}
    \label{fig:TZ_Ari_ClassicalActivity_TimeSeries_LS}
\end{figure}

We present the RV and activity indicator time series together with their GLS periodograms for $\epsilon$\,Eri in Fig.~\ref{fig:EpsEri_ClassicalActivity_TimeSeries_LS}. All time series exhibit significant power at the stellar rotation period and its first harmonic, both highlighted in purple. Similarly, Fig.~\ref{fig:TZ_Ari_ClassicalActivity_TimeSeries_LS} shows the corresponding data for TZ\,Ari. In this case, the RV, CON, and BIS time series display significant variability at the stellar rotation period as well as at its daily alias ($1-1/1.96 = 1/2.04~\mathrm{d}^{-1}$). In addition, we observe long-term variability in the RVs associated with the orbital motion of TZ\,Ari\,b. Since GLS models are sinusoidal, the planet's eccentricity shifts the highest peak slightly with respect to the literature value \citep{quirrenbach2022carmenes}. The CON time series exhibits a downwards trend, introducing excess power at low frequencies in the periodogram, while the FWHM shows a significant signal at 40.17\,d.

In Appendix~\ref{sec:appendix_table}, we list the significant periods and amplitudes retrieved via a pre-whitening procedure. This process involves iteratively identifying the most significant periodogram peak, fitting the RVs with a sinusoidal function of the same frequency as that peak, and subtracting it from the time series until no power crossing the 10$\%$ FAP threshold remains.

\subsection{Synthetic data} \label{sec:synthetic_data}

\subsubsection{StarSim modelling} \label{sec:starsim}

We generated synthetic time series data using the \texttt{StarSim}\footnote{\url{https://github.com/dbarochlopez/starsim}} code (\citealt{herrero2016modelling}; \citealt{rosich2020correcting}; Gomes et al., in prep.) in order to model the test stars. \texttt{StarSim} is a physically motivated stellar activity simulator that produces synthetic photometric and spectroscopic observables by modelling the inhomogeneous surface of a rotating star. The stellar disk is discretised into surface elements corresponding to three distinct components: quiet photosphere, cool spots, and bright faculae. Each component is represented using a high-resolution synthetic spectrum from the PHOENIX library \citep{husser2013new}, selected to match the appropriate temperature, surface gravity, and metallicity. The library also includes spectra computed at different limb angles to account for centre-to-limb variations across the stellar disk. Rather than synthesising full high-resolution spectra from the surface maps, the simulator computes the CCF corresponding to each surface element directly, resulting in a considerable reduction of the computing time. This CCF computation process explicitly emulates the \texttt{raccoon} pipeline, employing the identical masks and velocity sampling steps applied to the observational data.

We constructed tailored stellar models for each target star using constraints on stellar parameters from the literature (see Table~\ref{tab:stellar_params}). To estimate the contrast temperature difference between photosphere and spot and the convective blueshift, we adopted the empirical relations from \citet{herbst2021starspots} and \citet{liebing2021convective}. However, we allowed for wide uniform priors on these parameters rather than fixing them to a single best-fit value. This ensured that the relationships do not impose overly strong constraints on the simulated values, providing the NN with a diverse training set required to generalise well. 

We also simulated a large ensemble of stellar surface maps with active regions randomly distributed across the visible disk. Each active region is modelled as a circular spot, characterised by seven parameters: (1) appearance time, (2) lifetime, (3) colatitude, (4) longitude, (5) spot radius, (6) linear growth rate, and (7) linear decay rate. Surrounding each spot is a facular region, modelled as a concentric ring. The facular area is parametrized using a global facula-to-spot area ratio, $Q$, which is fixed for each simulation. As shown in previous studies \citep{lanza2003modelling,silva2010properties, dumusque2014soap, herrero2016modelling}, modelling circular spots surrounded by a facular corona successfully reproduces spot maps for high precision photometry of the Sun and other stars. 

However, several studies combining Solar Dynamics Observatory (SDO) images and magnetograms of the Sun with RVs have demonstrated that the Sun is facula-dominated, possessing an extensive network of standalone faculae that persists even during solar minimum when spots are absent \citep{collier2021separating, haywood2022unsigned}. Consequently, an independent treatment of spots and faculae is typically preferred to accurately simulate such networks \citep{zhao2025precise}. Given the high activity levels of our specific test targets (Sect.~\ref{sec:teststar}) and our objective to restrict the dimensionality of the parameter space, we keep the $Q$-factor approximation for this study. We note that this simplification is no longer present in the upcoming solar-benchmarked branch of the \texttt{StarSim} code (SunSim, Stucki et al. in prep.).

For simplicity, we assumed that each active region maintains a constant maximum size over most of its lifetime, with linear growth and decay timescales lasting a few days. Although more complex evolution laws for active regions could be implemented, the large number of simultaneously evolving regions, each with different lifetimes, locations, and appearance times, effectively captures the stochastic nature of surface inhomogeneities and approximates the effects of more realistic evolution scenarios.

We calibrate the simulations by adjusting the parameters of the active regions—specifically their number and size distributions—in order to reproduce the typical observed amplitudes and scatter in the spectroscopic time series. It is important to note that our objective was not to recreate the exact epoch-by-epoch RV pattern of the observations, nor did we perform a simultaneous fit (e.g. via Markov chain Monte Carlo, MCMC) of the spot contrast, convective blueshift, and active region geometries to the exact time series. Instead, our goal is to generate a robust training dataset that statistically matches the macroscopic properties of the observed stellar jitter, such as the overall root-mean-square (RMS) scatter, typical periodogram amplitudes, and the amplitude ratios between different CCF indicators. By maintaining a wide range for these parameters, we ensure the simulations retain sufficient generality to account for unexplained variability in the observed RV data, such as instrumental noise or unknown planetary signals. In Sect. \ref{sec:Synthetic gap} we discuss the current shortcomings of our simulations in matching some of these statistical metrics.

We simulated between 15 and 25 active regions for $\epsilon$\,Eri, while we used between 80 and 110 for TZ\,Ari. This number is significantly larger in the latter case due to its longer observational time baseline of over 1400\,d. Spot radii are drawn from a uniform distribution in the range of 3--6$^\circ$ for $\epsilon$\,Eri and 6--8$^\circ$ for TZ\,Ari. We also vary the $Q$ parameter, using values between 0 and 9 for $\epsilon$\,Eri. For TZ\,Ari, we fix $Q = 0$ as it is an M5 dwarf, where spot-induced variability dominates \citep{herrero2016modelling,baroch2020carmenes} and faculae have been found to be dark, rather than bright, in 3D magnetohydrodynamic simulations \citep{johnson2021forward,bhatia2025simulations}.

\subsubsection{Noise injection} \label{sec:Noise_injection}

In addition to modelling stellar activity using \texttt{StarSim}, we simulate the impact of photon noise on the CCF. To ensure our synthetic observations are realistic, we inject noise patterns that match the statistical characteristics of the actual observations. We quantify the noise budget for each target star by computing the ratio of the mean observational error of each spectroscopic data to the total root-mean-square (RMS) of its observed time series (Sect. \ref{sec:Coeff_errors}). This allows us to estimate the fractional contribution of noise to each time series and injects Gaussian noise according to this ratio. A summary of the estimated noise levels for each time series is provided in Appendix~\ref{sec:appendix_table}.

\section{Theoretical framework} \label{sec:Theoretical_framework}

When not considering the influence from instrumental effects, we assume that a spectroscopic observation $\boldsymbol{O}(\boldsymbol{v},t)$ in the exoplanet domain (e.g. spectrum, single line, CCF), depending on the vector of RV $\boldsymbol{v}$, and the time of observation $t$, can be influenced by the following two effects: (1) an orbiting planetary companion that induces a non-chromatic Doppler shift in $\boldsymbol{v}$, and which we denote in the following with $\boldsymbol{\epsilon}$, and (2) by stellar activity, which induces distortions on the spectral line shapes, including  apparent line shifts, and which we denote in the following with $\kappa$.

We can decompose any spectroscopic observation, $\boldsymbol{O}(\boldsymbol{v},t)$, as
\begin{equation}
\boldsymbol{O}(\boldsymbol{v},t) =  \sum^{N-1}_{n=0} a_n(t) \boldsymbol{B}_n(\boldsymbol{v}) + \boldsymbol{E}(\boldsymbol{v},t), \label{eq:First}
\end{equation} 
where $\boldsymbol{B_n}(\boldsymbol{v})$ is a set of functions, hereafter basis functions, parametrized by time-dependent coefficients $a_n(t)$. The term $\boldsymbol{E}(\boldsymbol{v},t)$ represents the residual uncertainty of the decomposition, and $N$ is the number of decomposition terms, where $N \leq N_v$. $N_v$ refers to the number of elements at which $\boldsymbol{v}$ is evaluated. Strictly speaking, no information is lost if $N = N_v$. However, our aim is to condense the original observation into a smaller number of components with minimal loss of physical information. In the following, we explain how to isolate an orthonormal component from $\boldsymbol{B}_n(\boldsymbol{v})$, which contains only information on the velocity shifts, from other components only containing information on the shape distortions.

\subsection{Isolating the line shift component} \label{sec:Isolating_shifts}

If we separate the contribution from Doppler shifts ($\boldsymbol{O_\epsilon}$) from the effects of stellar activity ($\boldsymbol{O_*}$),\footnote{We use the subscript $*$ to denote the total stellar activity because $\boldsymbol{O_*}$ encapsulates the entirety of the activity influence (both apparent line shifts and pure line shape distortions). The term $\kappa$, introduced in Eq.~\ref{eq:x_star}, specifically denotes only the bulk apparent line shift caused by this activity.} we can write
\begin{equation}
    \boldsymbol{O}(\boldsymbol{v},t)= 
    c_0(t) \boldsymbol{I}(\boldsymbol{v}) +
    \boldsymbol{O}_\epsilon(\boldsymbol{v},t) + \boldsymbol{O}_*(\boldsymbol{v},t) +\boldsymbol{E}(\boldsymbol{v},t), \label{eq:sum_eps_star}
\end{equation} 
where $\boldsymbol{I}(\boldsymbol{v})$ is a function of ones and $c_0(t)$ is a constant. In the case where the observation is only affected by $\epsilon$, this induces a shift:
\begin{equation}            \boldsymbol{v_\epsilon}(t)=\boldsymbol{v}+\epsilon(t)\boldsymbol{I}.
\label{eq:x_eps}
\end{equation}
Therefore, we can apply the Taylor expansion of $\boldsymbol{O_\epsilon}$ with respect to $\epsilon$ around $\boldsymbol{v}$: 
\begin{equation}
    \boldsymbol{O_\epsilon}(\boldsymbol{v_\epsilon}(t))= \boldsymbol{O}(\boldsymbol{v}+\epsilon(t)\boldsymbol{I}) = \boldsymbol{O_r}(\boldsymbol{v}) +\epsilon(t) \boldsymbol{O'_r}(\boldsymbol{v}) + \mathcal{O}(\epsilon^2), \label{eq:O_eps}
\end{equation} 
where $\boldsymbol{O_r}(\boldsymbol{v})$ is the undistorted rest frame, and $\boldsymbol{O'_r}(\boldsymbol{v})=\frac{d\boldsymbol{O_r}(\boldsymbol{v})}{d\boldsymbol{v}}$. In practical applications, this theoretical undistorted rest frame is approximated by our empirically derived time-averaged template CCF, $\boldsymbol{T}(\boldsymbol{v})$, as discussed in Sect. \ref{sec:template_construction}. Second-order expansion terms $\mathcal{O}(\epsilon^2)$ are negligible for the case of small $\epsilon$, typical of planetary companions. 

On the other hand, the stellar activity changes in $\boldsymbol{v}$ can be written in the most general way as
\begin{equation}            
\boldsymbol{v_*}(t)=\boldsymbol{v}+\kappa(t)\boldsymbol{I}+\boldsymbol{s}(t).
\label{eq:x_star}
\end{equation}
Here, $\boldsymbol{s}(t)$ is a vector with an unknown number of components describing pure line shape distortions. Since we do not know the functional dependence of these components on physical parameters, a multivariate Taylor expansion is not feasible. The only constraints we imposed on $\boldsymbol{s}(t)$ is that they are orthogonal to the line shift component, ensuring that $\kappa(t)$ captures the net line shift. In the most general way, we can then decompose the stellar activity part $\boldsymbol{O_*}(\boldsymbol{v},t)$ of the observation with a Taylor expansion with respect to $\kappa$ and capture the pure line shape distortions with $\boldsymbol{G_n}(\boldsymbol{v})$, hereafter the distortion basis function:
\begin{equation}
\boldsymbol{O_*}(\boldsymbol{v_*},t) = \boldsymbol{O_r}(\boldsymbol{v}) + \kappa(t) \boldsymbol{O'_r}(\boldsymbol{v}) + \sum^{N-1}_{n=2} g_n(t) \boldsymbol{G_n}(\boldsymbol{v}) \label{eq:O_star}
,\end{equation}  
where $g_n(t)$ are the time-dependent coefficients describing $\boldsymbol{G_n}(\boldsymbol{v})$. If we combine Eqs.~(\ref{eq:sum_eps_star}--\ref{eq:O_star}), we can write
\begin{equation}
\begin{split}
    \boldsymbol{O}(\boldsymbol{v_\epsilon+v_*},t) = c_0(t) \boldsymbol{I}(\boldsymbol{v}) + c_1(t) \boldsymbol{O_r}(\boldsymbol{v}) + l(t)\boldsymbol{O'_r}(\boldsymbol{v})
    \\+ \sum^{N-1}_{n=2} g_n(t)\boldsymbol{G_n}(\boldsymbol{v}) + \boldsymbol{E}(\boldsymbol{v},t), \label{eq:final}
\end{split}
\end{equation}

where $l(t)$ includes both the Doppler shift and the shift induced by the distorted line, i.e. $l(t) = \epsilon(t) + \kappa(t)$, and $c_1(t)$ is a scaling factor that multiplies the undistorted rest frame $\boldsymbol{O_r}(\boldsymbol{v})$, which in this case is equal to 1.\footnote{Strictly speaking, summing Eqs. \ref{eq:O_eps} and \ref{eq:O_star} results in a factor of 2 in front of the rest-frame profile $\boldsymbol{O_r}(\boldsymbol{v})$. We absorb this factor into $c_1(t)$ for notational simplicity, allowing $c_1(t)$ to naturally evaluate to $\approx 1$ and thereby retain its physical intuition as a direct tracer of the CCF amplitude changes.}

We thereby decomposed our observation $\boldsymbol{O}(\boldsymbol{v},t)$ into a basis function $\boldsymbol{B}$ that contains a line-shifted component captured by $\boldsymbol{O'_r}(\boldsymbol{v})$ and a distortion basis function, $\boldsymbol{G}(\boldsymbol{v})$. Comparing the above with Eq.~(\ref{eq:First}), we found
\begin{eqnarray}
    \boldsymbol{B}(\boldsymbol{v})&=&[\boldsymbol{I}(\boldsymbol{v}),\boldsymbol{O_r}(\boldsymbol{v}),\boldsymbol{O'_r}(\boldsymbol{v}),\boldsymbol{G_2}(\boldsymbol{v}),...,\boldsymbol{G_{N-1}}(\boldsymbol{v})],\label{eq:Bx_at}\\
    \boldsymbol{a}(t)&=&[c_0(t), c_1(t),l(t), g_2(t),..., g_{N-1}(t)].
\end{eqnarray} 
We note, that $\epsilon$ and $\kappa$ cannot be determined independently and are mathematically degenerate. This degeneracy can only be broken by relating the physical or statistical properties of the $c_0$, $c_1$ and $g_n$ coefficients to $\kappa$ (see Sect.~\ref{sec:convatt}). 

\subsection{Distortion basis functions} \label{sec:Basis functions}

For the general framework many different choices for the distortion basis functions $\boldsymbol{G_n}(\boldsymbol{v})$ can be used. PCA isolates the principal modes of variability of $\boldsymbol{O}(\boldsymbol{v},t)$. But this method does not distinguish between the different sources of variability (Doppler, activity and instrumental) nor does it ensure the orthogonality of its components to $\boldsymbol{O'_r}(\boldsymbol{v})$. PCA was applied in Doppler-free observations on an orthogonalised $\boldsymbol{O}(\boldsymbol{v},t)$ with respect to $\boldsymbol{O'_r}(\boldsymbol{v})$ in \citet{klein2024investigating}, and on the autocorrelation function of $\boldsymbol{O}(\boldsymbol{v},t)$, in \citet{collier2021separating}. An alternative option would be to use the Fourier basis, as proposed in \citet{zhao2020fiesta}. The basis consists of sine and cosine functions of different frequencies that are orthogonal to each other. However, since the basis is periodic over all $\boldsymbol{v}$, it puts equal weight on the central part of the CCF and in the wings, which are heavily affected by noise.

For a Gaussian-like observation $\boldsymbol{O}(\boldsymbol{v},t)$, such as a CCF or a specific spectral line, we propose using the Hermite basis as the distortion basis function, which is defined as
\begin{equation}
    \boldsymbol{G_n}(\boldsymbol{x}) = \frac{e^{-\boldsymbol{x}^2/2}}{\sqrt{2^n n! \sqrt{\pi}}} \boldsymbol{H_n}(\boldsymbol{x}), \quad
    \boldsymbol{H_n}(\boldsymbol{x}) = (-1)^{n} e^{\boldsymbol{x}^{2}} \frac{d^{n}}{d\boldsymbol{x}^{n}} e^{-\boldsymbol{x}^{2}},
\end{equation}
where $\boldsymbol{x} = (\boldsymbol{v} - \boldsymbol{v_0})/\sigma$ is the normalised velocity coordinate centred at $\boldsymbol{v_0}$, scaled by the width $\sigma$, and $\boldsymbol{H_n}(\boldsymbol{x})$ are the Hermite polynomials. These are defined by the recurrence relation
\begin{equation}
\boldsymbol{H_{n+1}}(\boldsymbol{x}) = 2\boldsymbol{x} \boldsymbol{H_n}(\boldsymbol{x}) - 2n \boldsymbol{H_{n-1}}(\boldsymbol{x}), \label{eq:hermite_recurrence}
\end{equation}
where $H_0(x) = 1$ and $H_1(x) = 2x$.

A Hermite basis can be understood as a weighted generalisation of the monomial basis $\boldsymbol{x}^n$, which is used to compute the central moments of a function. However, using pure monomials as basis functions leads to their divergence at large $|\boldsymbol{x}|$. By introducing a Gaussian weight function to the monomial basis function, the Hermite basis effectively suppresses these divergences and delivers an estimate of the orthonormal central moments of a Gaussian-like function $\boldsymbol{O}(\boldsymbol{v},t)$. Crucially, a Gaussian multiplied by Hermite polynomials forms a basis that is mathematically orthogonal and complete. This completeness is a primary motivation for our choice, as it guarantees that any smooth deviation or asymmetry in the line profile can be accurately represented.

In principle, the distortion basis functions can be constructed using the $n$-th order derivatives of the Gaussian fit to the observation $\boldsymbol{O}(\boldsymbol{v})$, with $n>2$, given that the $\boldsymbol{G_0}(\boldsymbol{x})$ and $\boldsymbol{G_1}(\boldsymbol{x})$ components are already accounted for in our general basis function $\boldsymbol{B}(\boldsymbol{v})$ by the $\boldsymbol{O_r}(\boldsymbol{v})$ and $\boldsymbol{O'_r}(\boldsymbol{v})$ components (Eq. \ref{eq:Bx_at}). For non-Gaussian profiles, such as M dwarfs similar to TZ\,Ari (Fig.~\ref{fig:CCF_basis_function_fit_EpsEri_TZ_Ari}), we apply a multi-Hermite basis. We model the reference $\boldsymbol{O_r}(\boldsymbol{v})$ by fitting a sum of two Gaussians, prioritising the simplest solution with the fewest free parameters. In the case of TZ\,Ari, the optimal fit combines a narrow positive Gaussian for the core with a broad negative Gaussian for the wings (Fig.~\ref{fig:CCF_basis_function_fit_EpsEri_TZ_Ari}).

\subsection{Calculation of coefficients} \label{sec:Gram-Schmidt}

We determined the optimal coefficients $a_n$ through a $\chi^2$ minimisation process:
\begin{equation}
    \chi^2 = \frac{1}{2}\langle\boldsymbol{O}(t)- \boldsymbol{a}(t)\boldsymbol{B}\rangle^2.
\end{equation} 
This process can be rewritten as  
\begin{equation}
    \chi^2 = \frac{1}{2}\langle \boldsymbol{O}(t)\rangle^2 +\frac{1}{2}\sum_{n=0}^{N}\sum_{n'=0}^{N}\left(a_n(t)a_{n'}(t)\langle      \boldsymbol{B_n},\boldsymbol{B_{n'}}\rangle-2a_n(t)\langle \boldsymbol{O}(t),\boldsymbol{B_{n}}\rangle\right).
\end{equation}  
If we orthonormalise $\boldsymbol{B_n}$, so $\langle B_n, B_{n'}\rangle=\delta_{nn'}$, this simplifies to  
\begin{equation}
    \chi^2 = \frac{1}{2}\langle \boldsymbol{O}(t)\rangle^2 +\frac{1}{2}\sum_{n=0}^{N}(\langle a_n^2\rangle-2a_n\langle \boldsymbol{O}(t),\boldsymbol{B_{n}}\rangle).
\end{equation}  
The coefficients that minimise $\chi^2$ are then obtained by solving  
\begin{equation} \label{eq:final_Chi2}
    \frac{d\chi^2}{da_n} = a_n(t) - \langle \boldsymbol{O}(t),\boldsymbol{B_n}\rangle=0
\end{equation}  
and hence can be directly determined from the basis function $\boldsymbol{B_n}$ and the observation $\boldsymbol{O}(t)$. The different $a_n(t)$ coefficients are independent from each other if $\boldsymbol{B}(\boldsymbol{v})$ is orthogonal. As explained in Sect. \ref{sec:Basis functions}, the Hermite basis is an orthogonal basis, but it may not be orthogonal to the other components of the basis $\boldsymbol{B}(\boldsymbol{v})$, namely $\boldsymbol{I}(\boldsymbol{v})$, $\boldsymbol{O_r}(\boldsymbol{v})$, or $\boldsymbol{O'_r}(\boldsymbol{v})$.

The orthogonality of $\boldsymbol{B}(\boldsymbol{v})$ can be achieved using the algebraic Gram-Schmidt process, which transforms any set of linearly independent functions into an orthogonal set. Given a non-orthogonal function $\boldsymbol{F_N}$, we orthogonalised it with respect to $\boldsymbol{B_n}$ using the iterative procedure  
\begin{eqnarray}
    \boldsymbol{f_{N}} = \boldsymbol{F_{N}} - \sum_{n=0}^{N} \langle \boldsymbol{B_n}, \boldsymbol{F_{N}} \rangle \, \boldsymbol{B_N},\\
    \boldsymbol{B_{N}} = \frac{\boldsymbol{f_{N}}}{\|\boldsymbol{f_{N}}\|},
    \label{eq:GramSchmidt}
\end{eqnarray} 
where $\boldsymbol{f_{N}}$ is the intermediate unnormalised function and $\|\cdot\|$ denotes its norm.

\subsection{Coefficient errors} \label{sec:Coeff_errors}

The uncertainties in the coefficients are obtained by propagating the errors from two independent sources: (1) the flux uncertainties of the observations, and (2) the uncertainties in the single- or multi-Gaussian fit to the template, from which the distortion basis functions are constructed. The total uncertainty in each coefficient can therefore be expressed as
\begin{equation}
    \delta a_n(t) = \sqrt{
    \sum_{i=1}^{N_v}
    \left[B_n(v_i)\,\delta O(v_i,t)\right]^2
    +\left[O(v_i,t)\,\delta B_n(v_i)\right]^2}.
\end{equation}

The uncertainty in the observed profile $\boldsymbol{O}(\boldsymbol{v},t)$ includes both photon noise and read-out noise (Appendix~\ref{sec:appendix_CCF_error}), as described in \citet{lafarga2020carmenes}. To compute realistic noise estimates per pixel, $\boldsymbol{O}(\boldsymbol{v},t)$ must be sampled at the physical pixel size of the detector. However, the Gram-Schmidt orthonormalisation used to construct the basis functions $\boldsymbol{B_n}$ makes an analytical expression for $\boldsymbol{\delta B_n}$ intractable. Instead, we propagated the uncertainties from the Gaussian-fit parameters into the orthonormalised basis by computing the formal covariance from the fit and propagating it through the coefficient definition
\begin{equation}
    \delta a_{n,\mathrm{basis}}^2(t) =\boldsymbol{J_n}^{T}\,\boldsymbol{\Sigma_p}\,\boldsymbol{J_n},
\end{equation}
where $\boldsymbol{J_n}$ is the Jacobian vector of partial derivatives of the coefficient $a_n(t)$ with respect to the set of fitted parameters $\boldsymbol{p}$, and $\boldsymbol{\Sigma_p}$ is the full covariance matrix of the fit. The Jacobian is defined as
\begin{equation}
    \boldsymbol{J_n} = \frac{\partial a_n(t)}{\partial \boldsymbol{p}}
    =\left[\frac{\partial a_n(t)}{\partial p_1},
    \frac{\partial a_n(t)}{\partial p_2},
    \ldots,\frac{\partial a_n(t)}{\partial p_k}\right],
\end{equation}
where $k$ is the number of fit parameters. Since these derivatives are not analytically tractable, we estimated them numerically using finite differences. For each parameter $p_k$, we computed
\begin{equation}
    \frac{\partial a_n(t)}{\partial p_k}
    \approx 
    \frac{a_n(t)\big|_{p_k + \Delta} - a_n(t)\big|_{p_k}}{\Delta},
\end{equation}
where $\Delta$ is a small perturbation to the parameter $p_k$.

\subsection{Explained variance ratio}\label{sec:Explained variance Ratio}

To evaluate the contribution of each coefficient $a_n$ to the reconstruction of the original time series, we compute the explained variance ratio (EVR), a metric analogous to that used in PCA. The EVR quantifies the fraction of the total variability in the full time series of $\boldsymbol{O}(\boldsymbol{v},t)$ captured by a given time series coefficient:
\begin{equation} \label{eq:explained_variance}
    EVR_n=\frac{\sum_{j=0}^{N_T}a_n(t)^2}{1/N_v\sum_{i=1}^{N_v}\sum_{j=0}^{N_T}\left[\boldsymbol{w}(v_i)\left(\boldsymbol{O}(v_i,t_j)-\langle \boldsymbol{O}(v_i)\rangle_t\right)\right]^2},
\end{equation}
where, in the most general case, a weight function $\boldsymbol{w}$ can be used for the observations. Further, $\langle\rangle_t$ is the expected value over $t$, and $N_T$ is the number of observations.

\section{Application of the method} \label{sec:Application_method}

Several considerations must be taken into account when applying the theoretical framework to real observational data. The observations shall be flux normalised before applying this framework using, for example, AFS \citep{xu2019modeling}, {\tt rassine} \citep{cretignier2020rassine}, or {\tt raccoon} \citep{lafarga2020carmenes}. If the continuum is not properly corrected to match the baseline flux level of $\boldsymbol{O_r}(\boldsymbol{v})$, this propagates to the computation of all $a_n(t)$, given that this continuum level is a multiplicative factor in Eq. (\ref{eq:First}). Moreover, real observational data are affected by noise, which can bias the estimate of the EVR. We adopt the Gaussian fit to $\boldsymbol{O_r}(\boldsymbol{v})$ as a weight function, so we can properly weigh down the wings of the CCF, which are predominantly dominated by noise.

\subsection{Template construction} \label{sec:template_construction}

As seen in Eq. (\ref{eq:final}), the decomposition is performed with respect to $\boldsymbol{O_r(\boldsymbol{v})}$, which represents the observation in an undistorted reference frame. This reference is also used to compute its first derivative, $\boldsymbol{O'_r}(\boldsymbol{v})$, in order to measure the line shift $l(t)$. In general, however, we do not have access to an undistorted reference observation $\boldsymbol{O_r}(\boldsymbol{v})$. We therefore decomposed our observations with respect to a template $\boldsymbol{T}(\boldsymbol{v})$ that is constructed by averaging all observations of the time series, $t$, as explained in Sect.~\ref{ssec:specdat}. We aim to reduce the influence of time-variable distortions and obtain a more stable representation of the stellar signal. The template provides a valid approximation to the undistorted reference profile required for the decomposition. Given that we use a template, generally, $c_1 \neq 0$, and it will estimate the changes in amplitude.

Consequently, we approximated $\boldsymbol{O'_r}(\boldsymbol{v})$ by the first derivative of the template $\boldsymbol{T}(\boldsymbol{v})$. We note that $\boldsymbol{T'}(\boldsymbol{v})$ was calculated from the first derivative of a cubic spline on $\boldsymbol{T}(\boldsymbol{v})$. The first element of the Gram-Schmidt process is $\boldsymbol{T'}(\boldsymbol{v})$, so we could ensure that all other elements of our basis function are orthogonal to line shifts.

\subsection{$\epsilon$\,Eri} \label{sec:Aplication_EpsEri}

\begin{figure}
    \centering
    \includegraphics[width=\columnwidth]{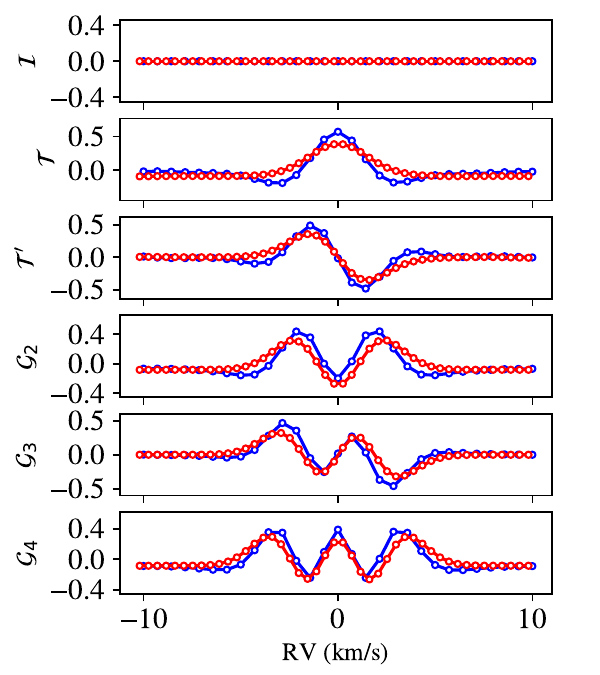}
    \caption{First six basis components for $\epsilon$\,Eri (Hermite, red) and TZ\,Ari (multi-Hermite, blue).}
    \label{fig:Reconstruction_basis_variance_TZ Ari_EpsEri_A}
\end{figure}

\begin{figure}
    \centering
    \includegraphics[width=\columnwidth]{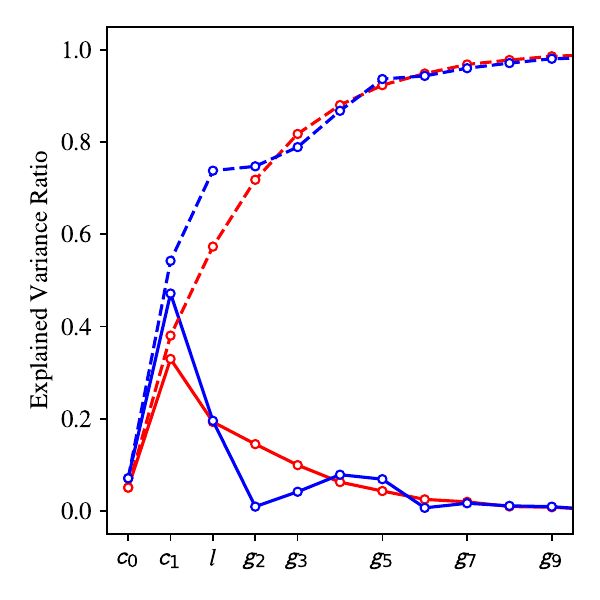}
    \caption{Explained variance ratio (solid) and cumulative variance (dashed) per component for $\epsilon$\,Eri (Hermite, red) and TZ\,Ari (multi-Hermite, blue).}
    \label{fig:Reconstruction_basis_variance_TZ Ari_EpsEri_B}
\end{figure}

As can be seen in the left panel of Fig.~\ref{fig:CCF_basis_function_fit_EpsEri_TZ_Ari}, the template of $\epsilon$\,Eri is in good agreement with a single Gaussian profile. We therefore adopted a single Hermite basis for its decomposition. In Fig.~\ref{fig:Reconstruction_basis_variance_TZ Ari_EpsEri_A} the first six components of the decomposition of the $\epsilon$\,Eri CCFs are shown in red, including the Hermite basis from $\boldsymbol{G_2}$ to $\boldsymbol{G_4}$. In Fig.~\ref{fig:Reconstruction_basis_variance_TZ Ari_EpsEri_B}, we show in red the individual and cumulative EVR per component. The individual EVR peaks at the $c_1$ component, connected to the template $\boldsymbol{T}$, and decreases monotonously as we go to higher order distortion coefficients. The first five coefficients explain 91.7\,\% of the variance, while 11 components account for 97.8\,\%. Notably, our method enables the quantification of the contribution to the variance by line shifts via the $l$ coefficient, which alone accounts for 19.3\,\% of the total variance.

\begin{figure*}[t!]
    \centering
    \includegraphics[width=0.95\textwidth]{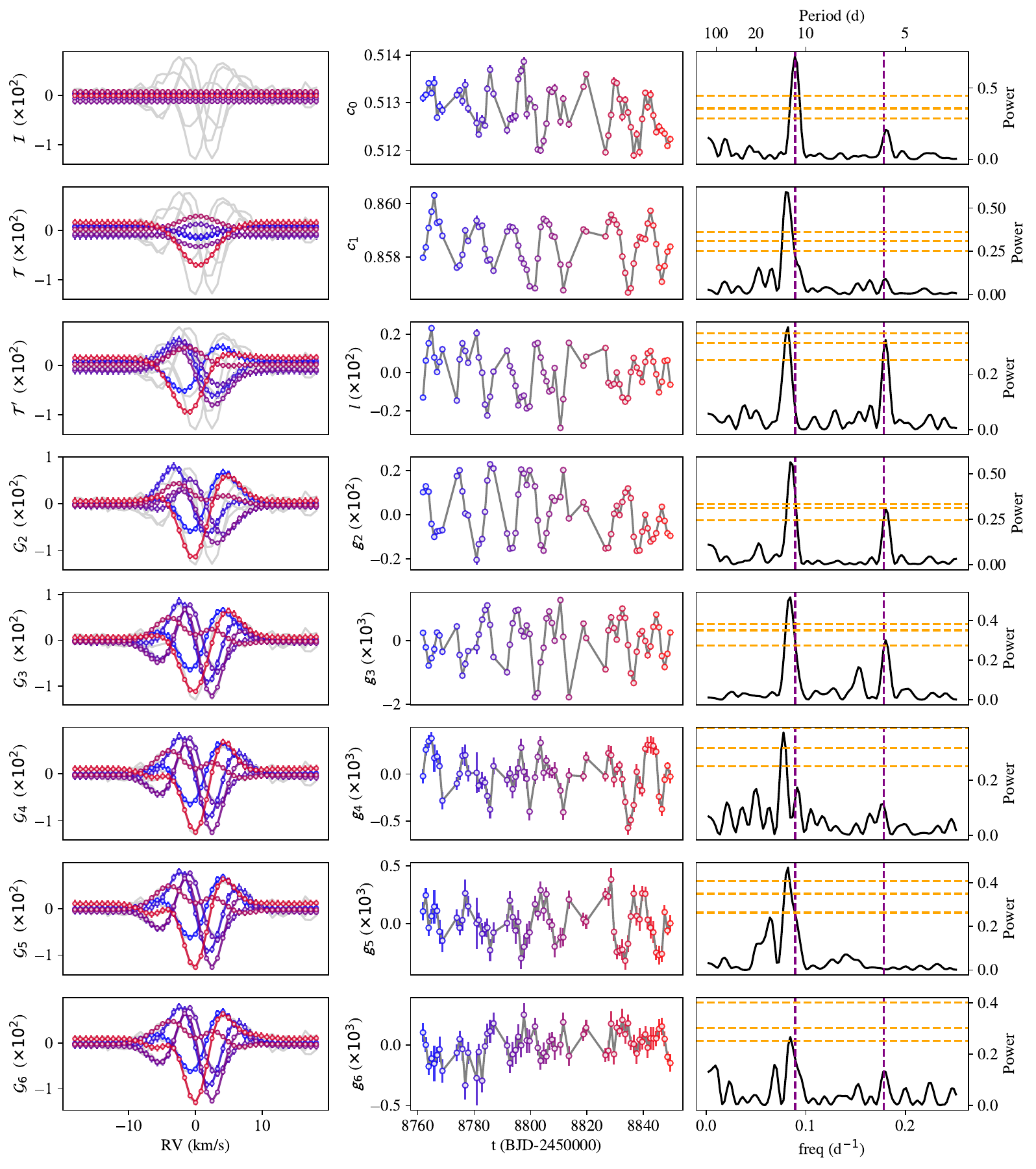}
    \caption{Cross-correlation function decomposition of $\epsilon$\,Eri with a Hermite basis. Left: Contribution of consecutive basis components to the differential CCF reconstruction. We displayed six illustrative epochs randomly selected to visualise the temporal variability of the line-shape distortions. The observed differential CCFs are shown in grey, while the reconstructed components are colour-coded according to the observation time. Middle: Time series of the corresponding coefficients. Right: GLS periodograms of the coefficients. The 0.1\,\%, 1\,\%, and 10\,\% FAP levels are indicated in orange, and the stellar rotation period and its first harmonic are marked in purple.}
    \label{fig:Reconstruction_timeseries_periodogram_EpsEri}
\end{figure*}

In Fig.~\ref{fig:Reconstruction_timeseries_periodogram_EpsEri} we show (i) the reconstruction of the differential (template-subtracted) CCFs at six random epochs by adding the contribution from the different basis function components (left column), (ii) the time series of the first eight coefficients (middle column), and (iii) their corresponding GLS periodograms, (right column). Coefficients up to $g_6$ trace the stellar rotation period, marked with purple lines, with $l$, $g_2$ and $g_3$ also showing sensitivity to the first harmonic at $\sim$5.6~d. The slight shifts in the exact periodogram peaks across different coefficient orders can be physically attributed to stellar differential rotation (as reported in \citealt{croll2006differential}) and active region evolution. We also note that the baseline offset term, $c_0$, also exhibits sensitivity to stellar rotation. While mathematically defined as a constant, $c_0$ naturally absorbs changes in the overall integrated area of the CCF, whether caused by imperfect flux normalisation or, as in this case, stellar activity driving variations in both the continuum level and line amplitude.

As expected from their lower EVRs, higher-order coefficients contribute less to the signal reconstruction and are increasingly affected by noise, therefore reducing their sensitivity to rotationally modulated stellar activity. For a summary of the general statistics of the distortion coefficients and the identified signals from a pre-whitening process we refer to Appendix~\ref{sec:appendix_table}.

Figure~\ref{fig:correlation_plot_EpsEri} shows the correlation between decomposition coefficients and classical CCF indicators. The strongest correlations are CON with $c_1$, RV with $l$, FWHM with $g_2$, and BIS with $g_3$, confirming that the decomposition recovers the classical CCF activity indicators. The strongest correlation between decomposition coefficients is between $l$ and $g_3$, and between $c_1$ and $g_2$. Beyond this, higher-order coefficients extend the sensitivity to distortions not captured by classical CCF activity indicators.

\subsection{TZ\,Ari} \label{sec:Application_TZAri}

Figure~\ref{fig:CCF_basis_function_fit_EpsEri_TZ_Ari} shows the template used to compute the basis functions. It displays the characteristic shape of cool M dwarfs \citep{lafarga2020carmenes}. As a result, a double Hermite basis is used to fit this template, consisting of a narrow positive-amplitude Gaussian for the core and a broader negative-amplitude Gaussian for the wings.  In Fig.~\ref{fig:Reconstruction_basis_variance_TZ Ari_EpsEri_A}, the first six components of the decomposition of the TZ\,Ari CCFs are depicted in blue, showing a distinct shape for the multi-Hermite case. Figure~\ref{fig:Reconstruction_basis_variance_TZ Ari_EpsEri_B} illustrates the EVRs for the multi-Hermite decomposition coefficients applied to TZ\,Ari. The EVR peaks at the $c_1$ coefficient, explaining 47.1\% of the EVR. The $l$ coefficient alone accounts for 19.6\,\% of the total variance. The $g_2$ coefficient contributes only marginally, while $g_5$ provides a comparatively large contribution to the variance. The cumulative EVR reaches 86.8\% at this coefficient, reaching the 97.1\% mark if we extend it up to the $g_{9}$ coefficient.

\begin{figure*}
    \centering
    \includegraphics[width=0.95\textwidth]{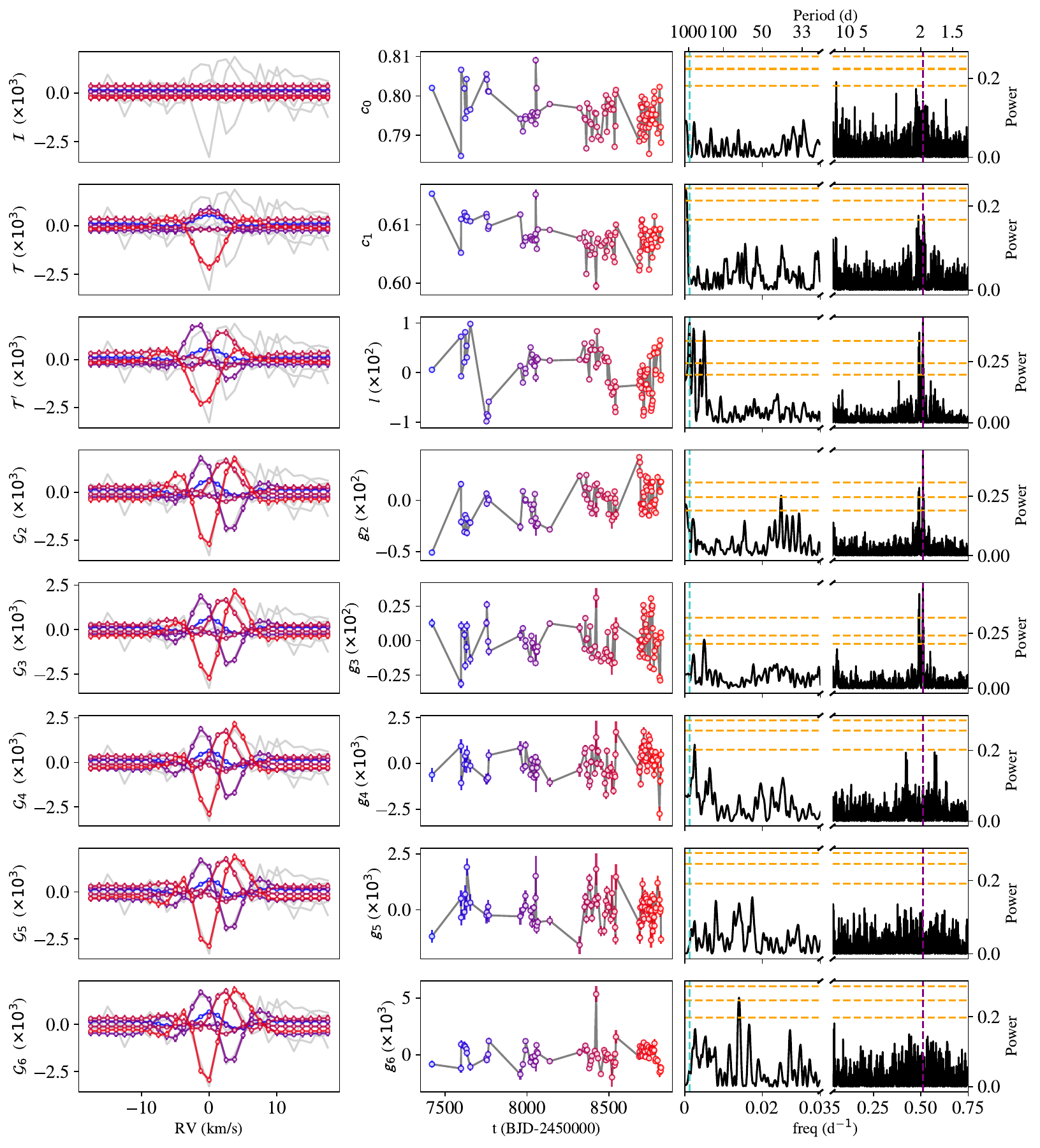}
    \caption{Same as Fig.~\ref{fig:Reconstruction_timeseries_periodogram_EpsEri} but for the multi-Hermite basis applied to TZ\,Ari. We also mark in light blue the planetary orbital period from the literature \citep[722.05\,d;][]{quirrenbach2022carmenes}.}
    \label{fig:Reconstruction_timeseries_periodogram_TZ Ari}
\end{figure*}

Figure~\ref{fig:Reconstruction_timeseries_periodogram_TZ Ari} shows the differential CCF reconstructions (left), coefficient time series (middle), and their corresponding GLS periodograms (right). The stellar rotation period (1.96\,d) and its daily alias are clearly recovered in the low-order coefficients; $c_1$, $l$, $g_2$ and $g_3$. Long-period signals are also observed in $c_1$, $l$, $g_2$, and $g_4$ (Appendix~\ref{sec:appendix_table}). The exoplanetary signal at 772.05\,d is significantly detected in the GLS periodogram of $l$ only, as expected. 

Figure~\ref{fig:correlation_plot_TZ Ari} displays correlations between decomposition coefficients and classical CCF activity indicators. As for the case of $\epsilon$\,Eri, the strongest correlations are found between CON and $c_1$, RV and $l$, FWHM and $g_2$, and BIS and $g_3$. However, the correlations between FWHM and $g_2$, and between BIS and $g_3$ are not as strong as in the $\epsilon$\,Eri case. The strongest correlation between decomposition coefficients is between $c_0$ and $c_1$, and between $l$ and $g_3$.

\section{Neural networks for stellar activity correction}\label{sec:convatt}

In this section we introduce the NN framework. The goal is to predict and remove the activity-induced RV shifts using the information from the line-shape distortion coefficients.

\subsection{Architectures} \label{sec:Architecture}

\begin{figure*}
    \sidecaption
    \includegraphics[width=12cm]{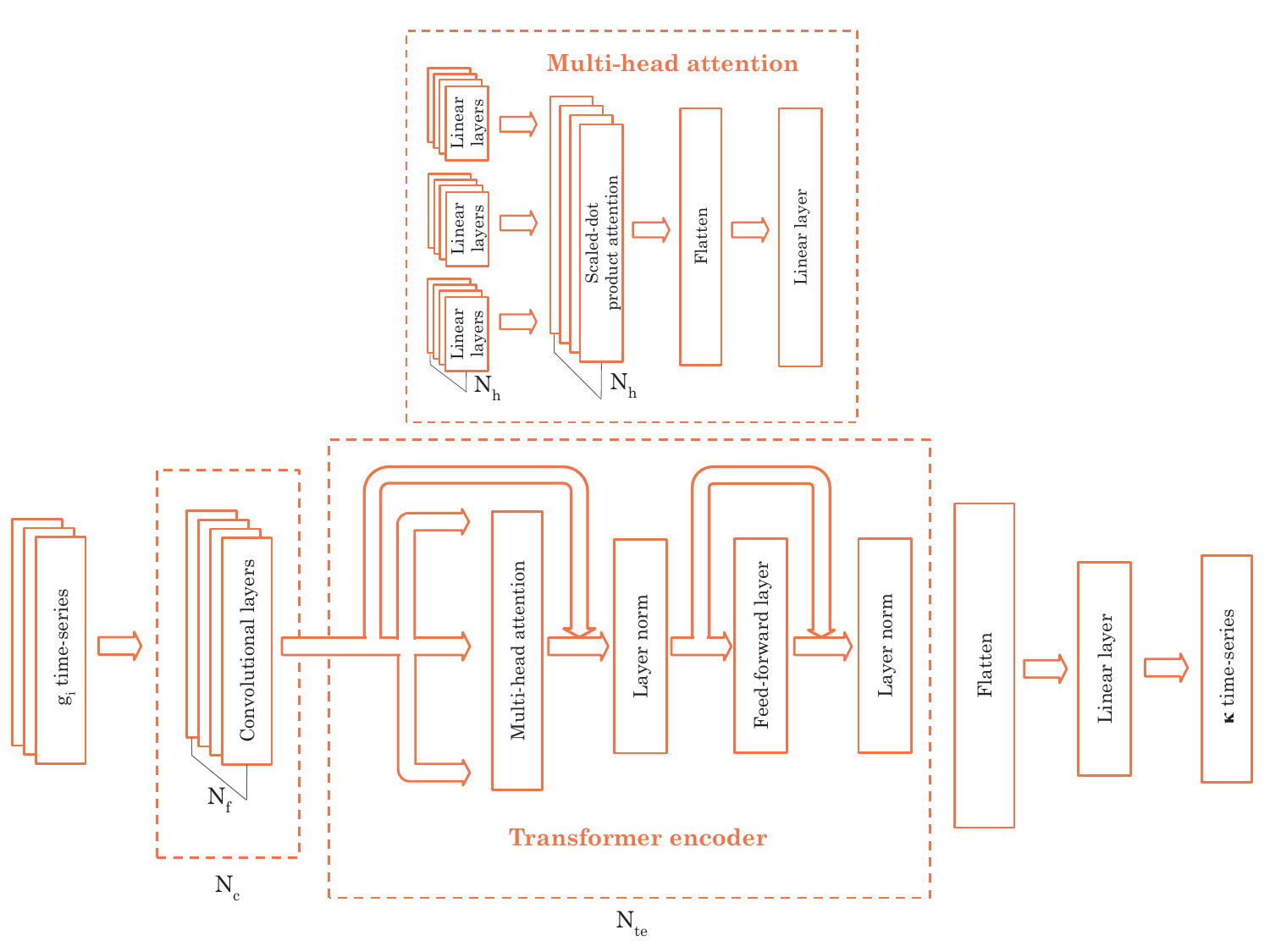}
    \caption{Schematic representation of \texttt{CANSTAR} (bottom), a convolutional attention network designed to predict the stellar activity contribution to line shifts, $\kappa(t)$, from the time series of distortion coefficients, $g_i(t)$. The input coefficients are treated as independent channels and processed through a series of one-dimensional convolutional layers that extract local temporal features. The convolutional outputs are then passed to a transformer encoder module, which consists of a multi-head self-attention part (top) with different `heads', where each one of them focuses on different temporal dependencies within the data, and their outputs are combined, normalised, and propagated through additional attention layers. Finally, the resulting features are flattened and projected through a linear layer to produce the predicted $\kappa(t)$ time series.}
    \label{fig:CANSTAR_architecture}
\end{figure*}

We have developed a deep-learning model that we refer to as Convolutional-Attention Network for STellar Activity Removal (\texttt{CANSTAR}). An overview of the \texttt{CANSTAR} architecture is provided in Fig.~\ref{fig:CANSTAR_architecture}. Several ($N_c$) convolutional layers apply $N_f$ localised filters to the input time series, which consist of the amplitude coefficient $c_1(t)$ and the distortion coefficients $g_i(t)$. We explicitly exclude the baseline offset term, $c_0(t)$, from the inputs because it is highly susceptible to flux normalisation issues and sensitive to noise sources not accounted for in our simulations. These filters extract short-term, high-frequency features in the input according to the filter size $N_s$. The convolutions are applied with appropriate padding and stride, ensuring that the output feature maps keep the same temporal dimension as the input. Each convolutional layer uses multiple filters, producing a set of feature maps that capture different local attributes of the data.

Following the convolutional layers, the extracted feature maps are passed into $N_{te}$ transformer encoder layers. These layers consist of a multi-headed self-attention mechanism followed by a position-wise feed-forward network, as defined in \citet{vaswani2017}. The self-attention mechanism captures dependencies across the full temporal domain, allowing the network to model both short- and long-term correlations in the data. The embedding dimension of the attention layer is matched to the number of convolutional filters, and $N_h$ attention heads are used to extract different temporal features. Finally, the output of the transformer block is flattened and passed through a linear projection layer, producing the predicted $\kappa(t)$ time series. This hybrid architecture follows principles similar to those proposed by \citet{liu2020convtransformer} and \citet{guo2022cmt}, who demonstrated the effectiveness of combining convolution and attention in the analysis of time series and image data.

We also tested a simpler fully connected network (FCN). This architecture processes only instantaneous snapshots of the line-shape distortion coefficients, $g_i$, without explicitly modelling their temporal evolution. The FCN architecture consists of a sequence $N_{ff}$ of feed forward layers. All input coefficients from a single observation are concatenated into a one-dimensional feature vector, which is fed into the network. The outputs of each layer are fully connected to the inputs of the next by $N_n$ number of neurons. The final layer of the FCN is a linear layer which predicts the final $\kappa$.

\subsection{Neural network training} \label{sec:Training}

Training is performed using synthetic datasets generated with the \texttt{StarSim} code (Sect.~\ref{sec:starsim}). For each target star, we generate a dataset of 500,000 time series simulations. The dataset is divided into training (80\,\%), validation (10\,\%), and test (10\,\%) sets. The training set is used to find the optimal parameters of the NN based on the minimisation of the Mean Squared Error (MSE) between the true labels, i.e. the $\kappa(t)$ component of $l(t)$, and the prediction by the network. We apply z-score normalisation to the input and output dataset, meaning we subtract the mean value of each simulated dataset and rescale them according to the standard deviation of the full dataset.

We optimised all network architectures using the Adam optimiser \citep{kingma2014adam} implemented in PyTorch \citep{paszke2019pytorch}. We selected as the final model the model just before the improvement in validation MSE plateaus while the training MSE continues to decrease significantly, ensuring we avoid the over-fitting regime. We reserved this test set for the evaluation of the performance of this final network on unseen data. We saved the residual relative error (RRE) from the final model, defined as the ratio of the residual RMS after correction to the original RMS \citep{perger2023machine}.

\subsection{Hyper-parameter choices}

Each architecture is defined by several hyper-parameters. We performed a systematic hyper-parameter search to determine these values. The tested and final selected configurations are summarised in the upper and lower parts of Table~\ref{tab:hyperparameters} for \texttt{CANSTAR} and the FCN, respectively. The final selection was based on minimising the validation set MSE, prioritising the simplest effective choices.

\begin{table}[h!]
\caption{Tested and adopted hyper-parameters for CANSTAR (top) and FCN (bottom).}
\centering
\tiny
\begin{tabular}{llll}
\hline \hline
Network & Hyper-parameter &  Priors & Result \\
\hline
\multirow{5}{*}{CANSTAR} & $N_s$ &  1,3,5,7 & 3 \\
 & $N_f$ &  4,8,16,32,64,128,256 & 128\\
 & $N_c$ &  1,2,3,4,5 & 4  \\
 & $N_h$ &  1,2,4,8,16,32 & 2 \\
 & $N_{te}$ &  1,2,3,4,5 & 4  \\
\hline
\multirow{2}{*}{FCN} & $N_n$ &  256,512,1024,2048 & 1024\\
 & $N_{ff}$ &  2,3,4,5,6,7 & 6  \\
\hline 
\end{tabular}
\label{tab:hyperparameters}
\end{table}

\section{Results} \label{sec:Results}

We trained separate \texttt{CANSTAR} models for each target star ($\epsilon$\,Eri and TZ\,Ari) for different noise-level regimes (noiseless and similar to the observations; see Sect.~\ref{sec:Noise_injection}) and including different numbers of distortion coefficients. The aim with this training strategy was to evaluate the ability of \texttt{CANSTAR} to mitigate stellar activity as a function of the target star and S/N of the observations. In order to test the capability of modelling stellar activity by the different distortion coefficients, we iteratively trained \texttt{CANSTAR} networks using more coefficients as inputs, following the natural order of the coefficients from the Hermite and multi-Hermite formalism. We repeated this same training strategy for the FCN.

After the training phase, we selected the best-performing network trained on noise similar to the observations, defined as the model that minimises the RRE while requiring a smaller number of distortion coefficients. To quantify the stability of the solution, we trained five independent instances of this architecture. Although the variance in the predictions on the synthetic test set is negligible, it can have an effect when predicting on the real observational data. To account for this, we applied the full ensemble of five networks to the data. Furthermore, to propagate the observational uncertainties, we generated 1000 Monte Carlo realisations of the input distortion coefficients, perturbed according to their measurement errors. Each of the five networks generates predictions for all 1000 realisations. We adopted the median of this combined distribution as the final activity model and used the standard deviation as the total uncertainty.

\subsection{Simulated data} \label{sec:Results simulations}

\begin{figure}
    \centering
    \includegraphics[width=\linewidth]{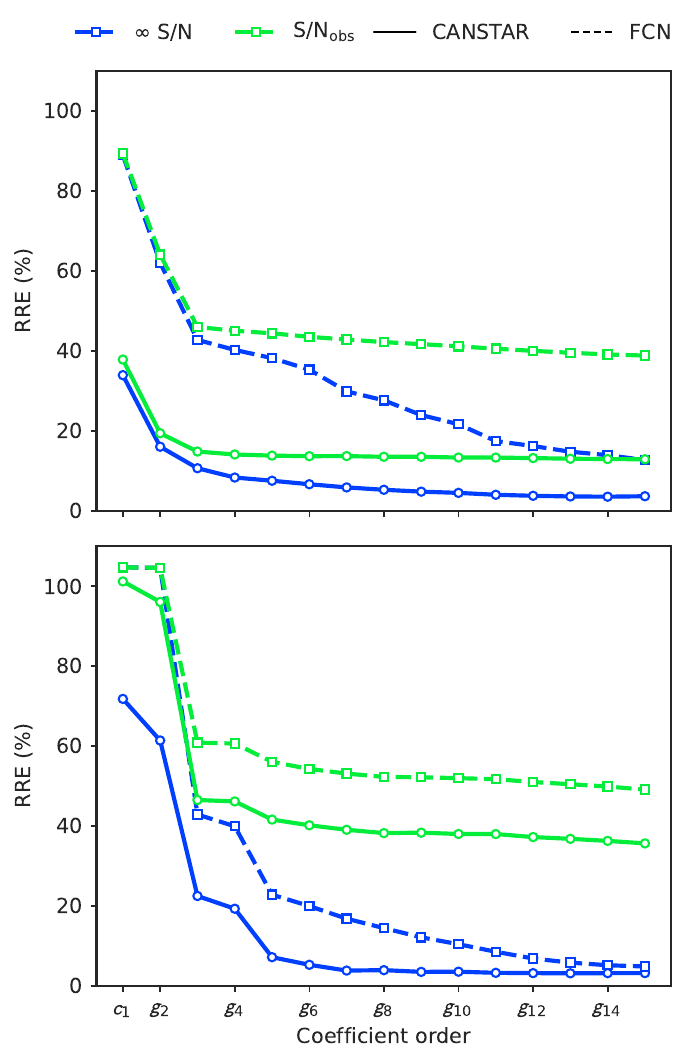}
    \caption{Residual relative error of $\epsilon$\,Eri (top) and TZ\,Ari (bottom) \texttt{StarSim} data after correction with \texttt{CANSTAR} (solid line with circle marker) and an FCN (dashed line with square marker) for the noiseless case (blue) and for the case with noise equivalent to the observations (green).}
    \label{fig:Performance_levels_noise}
\end{figure}

We show the performance on the simulated test set of \texttt{CANSTAR} and FCN in  Fig.~\ref{fig:Performance_levels_noise}. In the noiseless case, \texttt{CANSTAR} achieves an RRE of 3.7\,\% for $\epsilon$\,Eri and 3.3\,\% for TZ\,Ari, when trained with 15 coefficients. Notably, the performance already saturates at 7 coefficients, reaching an RRE 5.9\,\% for $\epsilon$\,Eri and 3.9\,\% for TZ\,Ari. By contrast, the FCN achieves an RRE of 12.7\,\% for $\epsilon$\,Eri and 4.9\,\% for TZ\,Ari after training with 15 coefficients. \texttt{CANSTAR} outperforms the FCN in RRE reduction, requiring fewer coefficients to reach the same level of activity correction.

With noise matched to the observed RV uncertainties, \texttt{CANSTAR} reaches an RRE of 12.9\,\% for $\epsilon$\,Eri and 35.7\,\% for TZ\,Ari, when trained with 15 coefficients. On the other hand, the FCN network reaches an RRE of 38.8\,\% for $\epsilon$\,Eri and 49.1\,\% for TZ\,Ari. In both regimes, the RRE reduction plateaus at a smaller number of coefficients than in the noiseless case, highlighting the fact that higher-order coefficients are increasingly dominated by noise and contribute little to the correction.

There is also a difference in the performance of the network between the two target stars. For $\epsilon$\,Eri, the most significant coefficients when correcting for activity are the $c_1$, $g_2$ and $g_3$ coefficient. In the case of TZ\,Ari, $c_1$ and $g_2$ are not as correlated to the line shift component as $g_3$, which is the most informative for training both \texttt{CANSTAR} and the FCN network. We also note that we reached a lower overall RRE reduction for the cases of $\epsilon$\,Eri with observational noise when compared to TZ\,Ari. This is because the former is a much brighter star with significantly higher S/N observations.

\subsection{$\epsilon$\,Eri} \label{sec:Results epsilon eri}

For $\epsilon$\,Eri, we selected the network trained on the first five distortion coefficients as the best-performing configuration. We applied z-score normalisation to the real observed datasets in order to minimise the discrepancy in the scale between observed and synthetic datasets (Appendix~\ref{sec:appendix_scale_coeffs}). We explored the inclusion of a linear scaling parameter to fine-tune the amplitude of the predictions given the loss of the absolute scale due to this. The scaling parameter converged to a value of $\alpha = 0.9$. This value is close to unity, as expected, given that the RV variability of $\epsilon$\,Eri is dominated by stellar activity on these timescales. Therefore, the $z$-score normalisation used during training accurately reflects the scale of the observed activity signal (Appendix~\ref{sec:appendix_scale_coeffs}).

\begin{figure*}
    \centering
    \includegraphics[width=\linewidth]{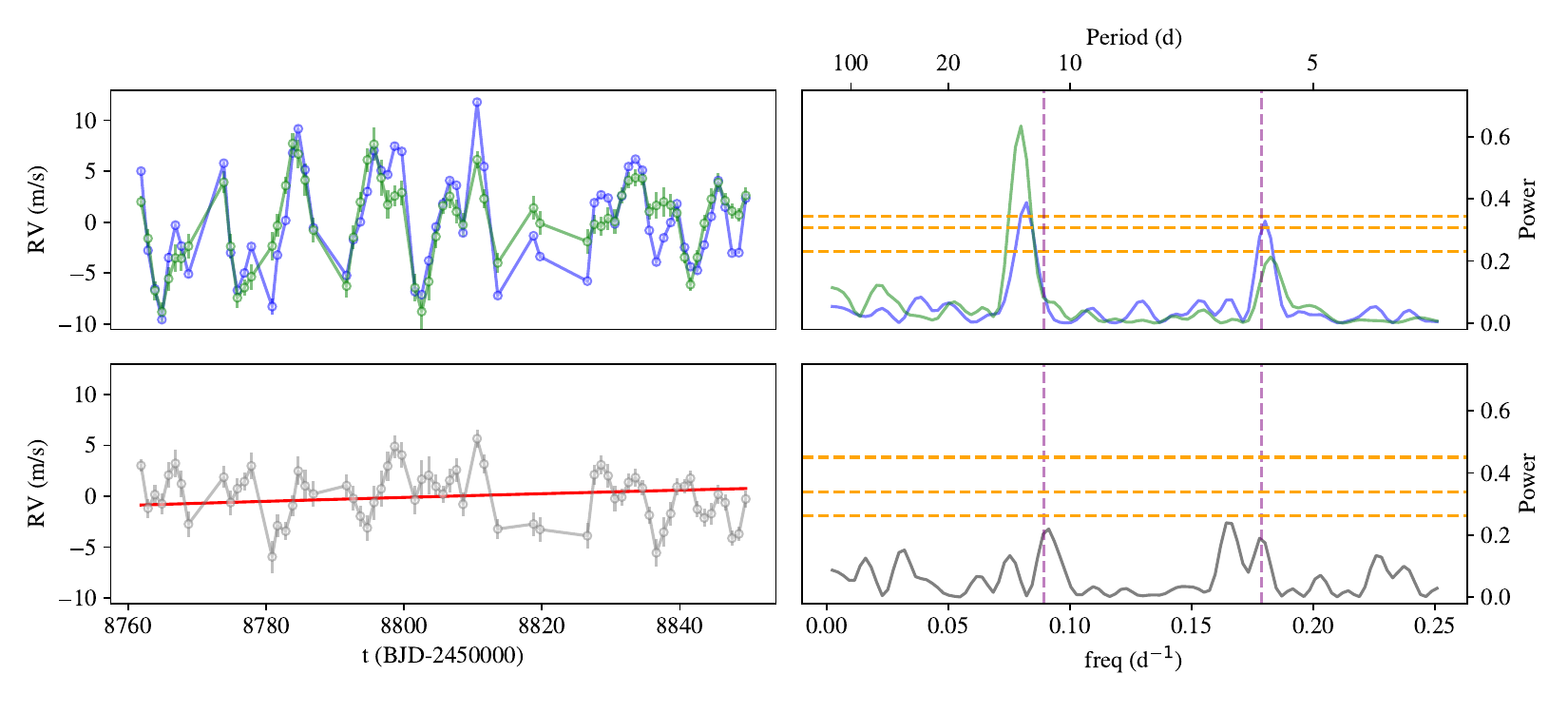}
    \caption{Top: Radial velocity time series (left) and GLS periodogram (right) of the HARPS $\epsilon$\,Eri data (blue) compared to the stellar activity signal predicted by \texttt{CANSTAR} (green). Bottom: Residuals after subtracting the prediction (grey), with their corresponding periodogram (right). The 10\,\%, 1\,\%, and 0.1\,\% FAP levels are indicated by dashed orange lines and the stellar rotation period (11.2\,d), and its first harmonic (5.6\,d) is marked in purple. The residuals are compatible with the $\epsilon$\,Eri~b solution \citep[red;][]{thompson2025revised}}
    \label{fig:correction_epseri}
\end{figure*}

\texttt{CANSTAR}'s correction for $\epsilon$\,Eri is shown in Fig.~\ref{fig:correction_epseri}. The predicted RV time series closely matches the observed data, and the GLS periodogram reveals that the model effectively captures the dominant stellar rotation signal and its first harmonic. After subtracting the predicted $\kappa$ component from the observations, we obtain an RRE of 52.5\,\%, resulting in a final RMS of $2.46\,\mathrm{m\,s^{-1}}$ (Table~\ref{tab:statistics_CANSTAR}). The power of signals associated with stellar activity, which peak at 12.53\,d and 6.26\,d, is significantly diminished, decreasing below the 10\,\% FAP level. No signal remains in the periodogram above the 10\,\% FAP level. The residuals are in agreement with the $\epsilon$\,Eri~b solution presented in \citet{thompson2025revised}, which just produces a trend for the narrow observing window.

\begin{figure*}
    \centering
    \includegraphics[width=\linewidth]{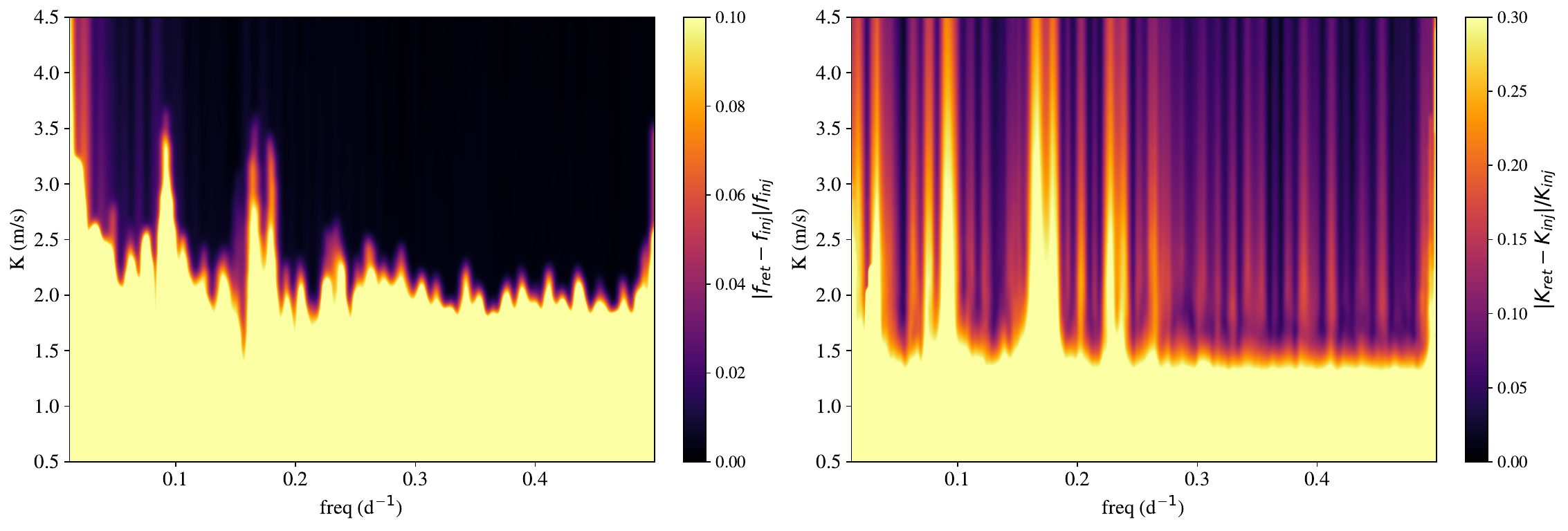}
    \caption{Detection limits in the $\epsilon$\,Eri residuals after applying \texttt{CANSTAR}'s correction, shown as a function of the difference between injected and retrieved frequencies (left) and semi-amplitudes (right) for the different injected frequencies and semi-amplitudes sinusoidal signals.}
    \label{fig:detection_limits}
\end{figure*}

We injected sinusoidal signals directly into the RV time series to quantify the detection limits at different frequency ranges. We injected sinusoidal signals with frequencies ranging from 0.5 to $1/T_{base}$ (where $T_{base}$ is the time baseline of the observations), in steps of $1/5T_{base}$. The semi-amplitude grid spans from 0.5\,m\,s$^{-1}$ to the RMS of the observations, 4.5\,m\,s$^{-1}$, in steps of 0.1\,m\,s$^{-1}$. For each combination of frequency and semi-amplitude, we injected ten evenly spaced phases from 0 to $2\pi$. The phase-averaged results are illustrated in Fig.~\ref{fig:detection_limits} as heat maps showing the relative difference between the injected and recovered semi-amplitudes ($K$) and frequencies for the \texttt{CANSTAR} activity-corrected residuals. We defined the detection limit as the minimum injected amplitude required to recover the semi-amplitude with a specific precision (e.g. 10\,\% or 30\,\%), provided the frequency is also retrieved within a 10\,\% relative error. Using this criterion, the detection limits are $2.51 \pm 0.43$\,m\,s$^{-1}$ for a 10\,\% precision threshold and $1.99 \pm 0.41$\,m\,s$^{-1}$ for a 30\,\% threshold. As expected, the recovery of planetary signals becomes more challenging around the stellar rotation period and its harmonics due to the inherent degeneracy between Keplerian and activity-induced variability. We also show in Appendix~\ref{sec:injection_retrieval_appendix} the detection limits for the null hypothesis case. The detection limits are $3.99 \pm 0.18$\,m\,s$^{-1}$ (10\,\% precision) and $3.87 \pm 0.18$\,m\,s$^{-1}$ (30\,\% precision).

\subsection{TZ\,Ari} \label{sec:Results TZ Ari}

For TZ\,Ari, we similarly select the network trained on the first three order coefficients as the best-performing configuration. Unlike $\epsilon$\,Eri, the RV variability of TZ\,Ari is dominated by the high-amplitude Keplerian signal of the planet \citep[$K=21.11$\,m\,s$^{-1}$;][]{quirrenbach2022carmenes}. Consequently, we need to perform a joint fit for the Keplerian solution of the planet together with the \texttt{CANSTAR} activity correction. We use the MCMC sampler emcee \citep{foreman2013emcee}, simultaneously solving for the orbital parameters of TZ\,Ari~b and the optimal value of the scaling parameter of the \texttt{CANSTAR} activity correction, $\alpha$. For the MCMC parameter optimisation, we employed 700 walkers, each run for 32,000 steps to ensure convergence of the chains. The first 2,000 steps were discarded as burn-in. The adopted priors and the resulting optimised parameter values are listed in Table~\ref{tab:posterior_planet}, together with the Keplerian solution we retrieve if we do not apply the \texttt{CANSTAR} activity correction. We also compare these results to a GP framework using a simple harmonic oscillator (SHO) kernel, reproducing the strategy of \citet{quirrenbach2022carmenes} but restricting their multi-instrument dataset to CARMENES only (Appendix~\ref{sec:appendix_GP}).

\begin{figure*}
    \centering
    \includegraphics[width=\textwidth]{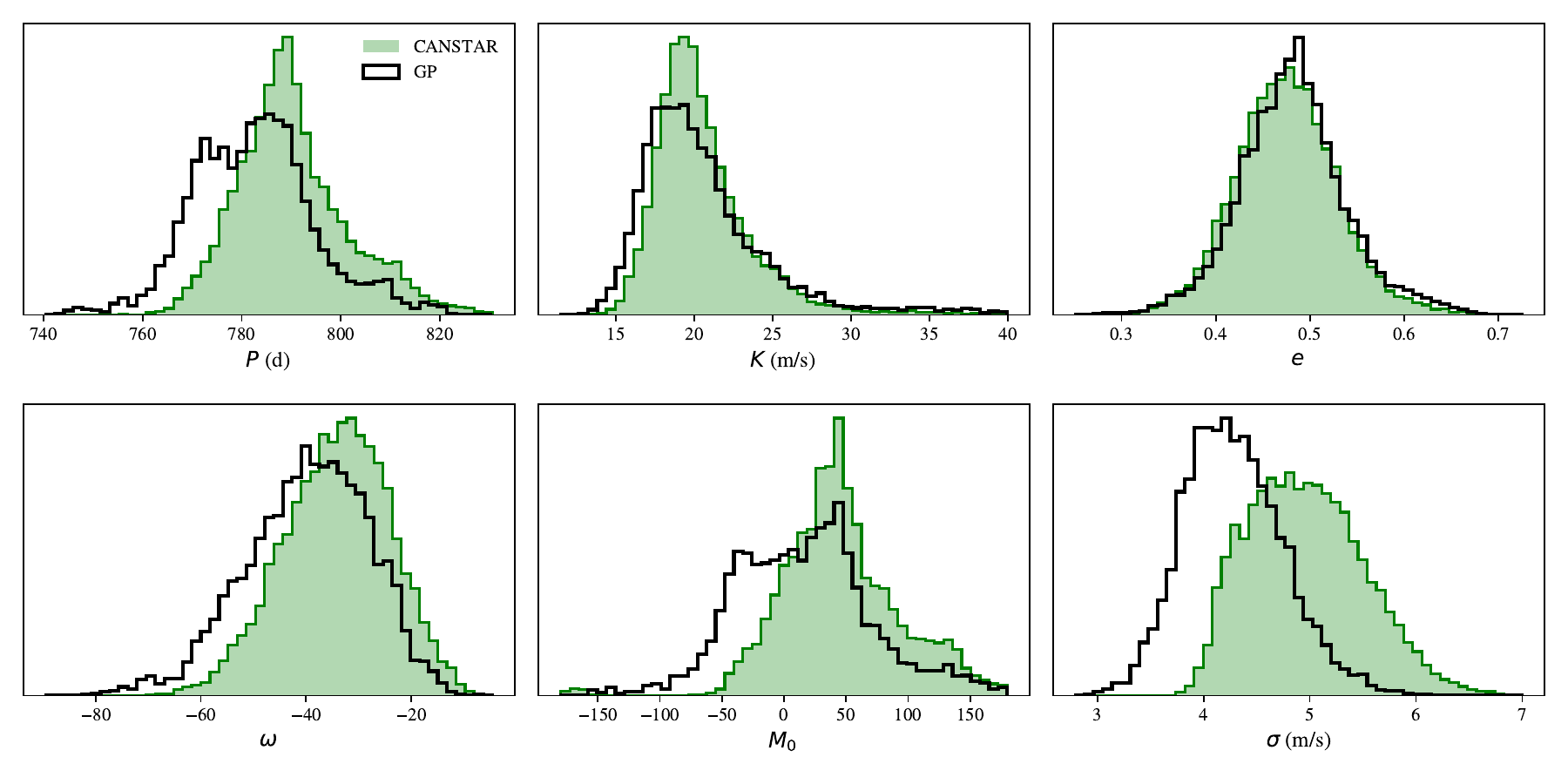}
    \caption{Histograms showing the posterior distribution of the shared parameters between the \texttt{CANSTAR} and GP correction methods used to model the TZ\,Ari RV data.}
    \label{fig:comparison_GP}
\end{figure*}

We compare in Fig.~\ref{fig:comparison_GP} the posterior distribution of the common parameters between the \texttt{CANSTAR} and GP model. \texttt{CANSTAR} activity correction allows for a better determination of the period and semi-amplitude of the planet, compared to the GP modelling. The GP solution outperforms \texttt{CANSTAR} in reducing the RV scatter, as illustrated by its lower fitted jitter term, which captures additional noise sources not captured by the model. This is expected given the strong flexibility of GPs at filtering white noise and being prone to overfitting \citep{blunt2023overfitting}. The posterior distributions of the \texttt{CANSTAR} + Keplerian model parameters are shown as corner plots in Appendix~\ref{sec:appendix_posteriors}.

\begin{table}
    \caption{Radial velocity statistics before and after \texttt{CANSTAR} correction.}
    \label{tab:rv_stats}
    \centering
    \begin{tabular}{l c c}
    \hline\hline
    Statistic [m s$^{-1}$] & $\epsilon$\,Eri & TZ\,Ari \\
    \hline
    RMS & & \\
    Original ($\text{RMS}_{\text{pre}}$) & 4.69 &  10.79\\
    Corrected ($\text{RMS}_{\text{post}}$) & 2.46 & 6.71 \\
    RRE & 52.5\,\% & 62.4\,\% \\
    
    \noalign{\smallskip}
    MAD & & \\
    Original $\text{MAD}_{\text{pre}}$ & 3.92 & 8.85 \\
    Corrected $\text{MAD}_{\text{post}}$ & 2.00 & 5.02 \\
    rMAD & 51.1\,\% & 57.4\,\% \\
    
    \noalign{\smallskip}
    Uncertainty Budget & & \\
    Photon Noise ($\sigma_{\text{phot}}$) & 0.29 & 2.44 \\
    Model Uncertainty ($\sigma_{\text{model}}$) & 1.07 & 4.49 \\
    Final Error $\left(\sqrt{\sigma_{\text{phot}}^2 + \sigma_{\text{model}}^2}~\right)$ & 1.11 & 5.19 \\
    \hline
    \end{tabular}
    \tablefoot{RMS: root mean square; RRE: residual relative error; MAD: median absolute deviation; rMAD: relative median absolute deviation. The original values for TZ\,Ari are given after subtracting the fitted Keplerian model.}
    \label{tab:statistics_CANSTAR}
\end{table}

{\renewcommand{\arraystretch}{1.6}
\begin{table*}
\caption{Comparison of orbital parameters for TZ Ari b.}
\centering
\begin{tabular}{l c c c c}
\hline
Parameter 
& Prior 
& Keplerian 
& \texttt{CANSTAR} 
& GP \\
\hline

$P$ [d] 
& $\mathcal{U}(650, 850)$
& $808.28^{+21.37}_{-17.57}$
& $789.92^{+12.17}_{-9.27}$
& $780.61^{+11.70}_{-11.66}$ \\

$K$ [m\,s$^{-1}$] 
& $\mathcal{U}(0, 40)$
& $22.09^{+3.82}_{-2.97}$
& $19.89^{+3.13}_{-1.98}$
& $19.91^{+4.08}_{-2.59}$ \\

$e$ 
& $\mathcal{U}(-1, 1)$ \tablefootmark{a}
& $0.55^{+0.07}_{-0.07}$
& $0.47^{+0.05}_{-0.05}$
& $0.48^{+0.06}_{-0.05}$ \\

$\omega$ [deg] 
& $\mathcal{U}(-1, 1)$ \tablefootmark{a}
& $-46.58^{+12.69}_{-13.74}$
& $-33.63^{+9.45}_{-10.86}$
& $-38.16^{+10.97}_{-12.35}$ \\

$M_{0}$ [deg] 
& $\mathcal{U}(0, 360)$
& $122.30^{+78.70}_{-70.01}$
& $42.31^{+49.63}_{-39.45}$
& $8.99^{+49.86}_{-50.68}$ \\

$\alpha$ [m\,s$^{-1}$]
& $\mathcal{U}(0, 30)$
& ---
& $12.60^{+0.89}_{-0.98}$
& --- \\

$S_{GP}$
& $\mathcal{N}(0.01, 0.005)$
& ---
& ---
& $0.0145^{+0.0040}_{-0.0036}$ \\

$Q_{GP}$
& $\mathcal{U}(5, 4473)$ \tablefootmark{b}
& ---
& ---
& $2347.84^{+1353.13}_{-1099.21}$ \\

$\omega_{GP}$
& $\mathcal{N}(3.2, 0.01)$
& ---
& ---
& $3.2108^{+0.0015}_{-0.0016}$ \\

$\sigma$ [m\,s$^{-1}$]
& $\mathcal{LU}(0.1, 30)$
& $9.87^{+0.92}_{-0.82}$
& $5.31^{+0.55}_{-0.47}$
& $4.31^{+0.55}_{-0.46}$ \\

$\gamma$ [m\,s$^{-1}$]
& $\mathcal{U}(-10, 10)$
& $-1.48^{+1.27}_{-1.25}$
& $-0.88^{+0.88}_{-0.79}$
& $-2.66^{+0.94}_{-0.76}$ \\
\hline
\end{tabular}
\tablefoot{
Values derived for CARMENES data modelling with the Keplerian alone or combining the fit with \texttt{CANSTAR} or with a GP fit. The prior column lists the probability distributions used in our MCMC analysis, where $\mathcal{U}(a,b)$ denotes a uniform distribution and $\mathcal{LU}(a,b)$ a log-uniform distribution between $a$ and $b$. \tablefoottext{a}{Prior applied to $\sqrt{e} \cos \omega$ and $\sqrt{e} \sin \omega$.} \tablefoottext{b}{Upper limit prior so we sample lifetimes up to two times the time baseline of the observations.}
}
\label{tab:posterior_planet}
\end{table*}}

\begin{figure*}
    \centering
    \includegraphics[width=\textwidth]{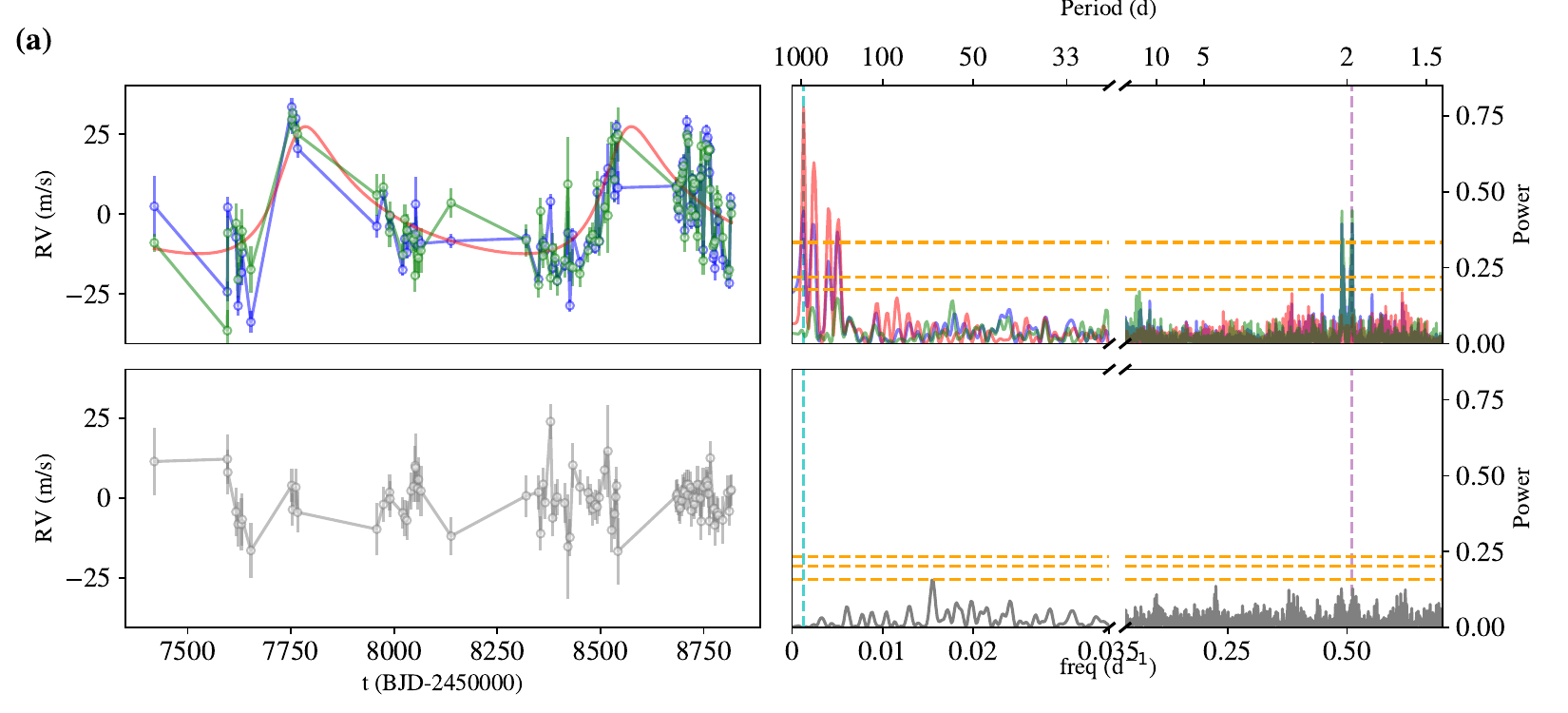}
    \includegraphics[width=\textwidth]{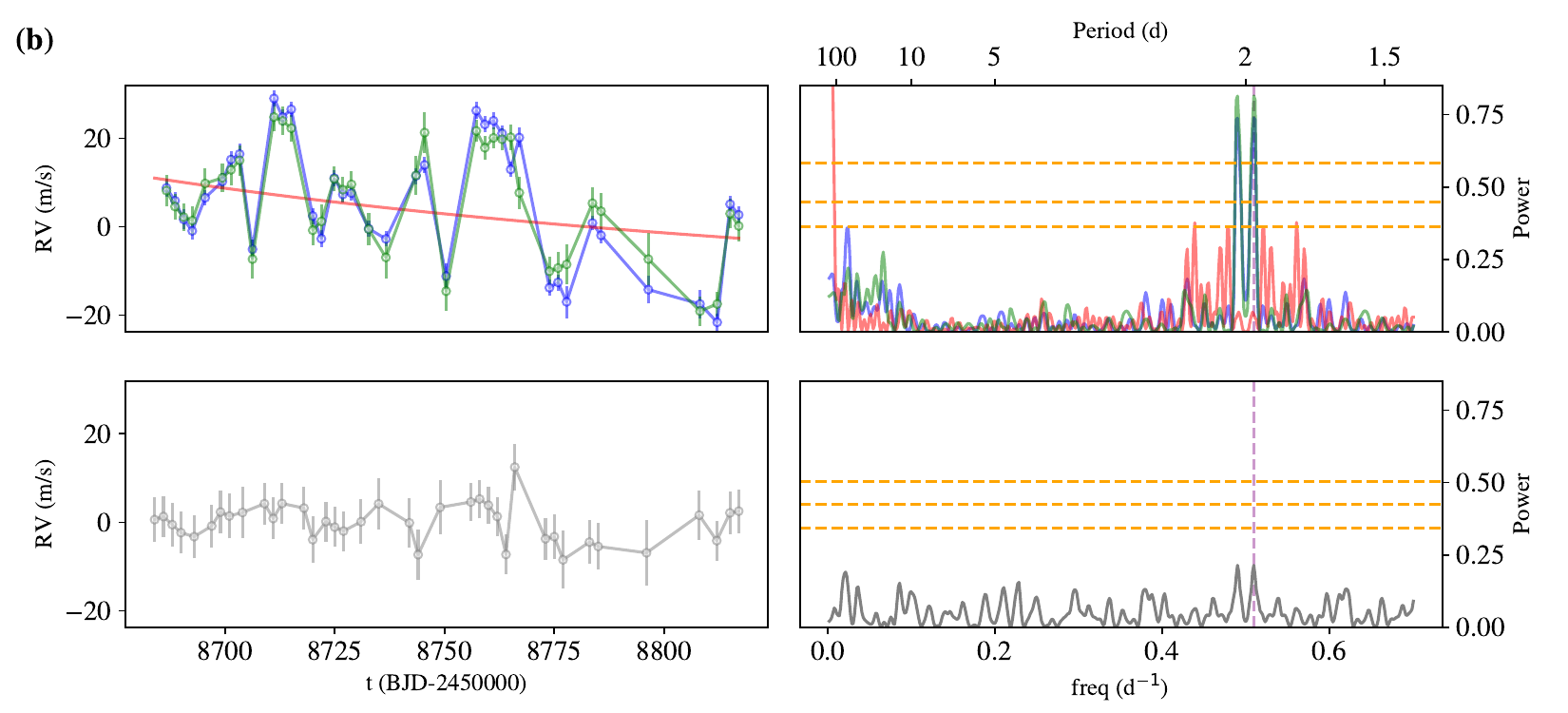}
    \caption{Radial velocity time series and GLS periodograms for TZ\,Ari showing \textbf{(a)} the complete multi-season dataset and \textbf{(b)} a detailed view of the final observing season. For each panel, the top row displays the CARMENES data (blue), the \texttt{CANSTAR} prediction (green), and the Keplerian solution (red). In the time series, the green curve includes the Keplerian signal to illustrate the full fit, whereas in the periodogram it represents the activity model alone. The bottom row shows the residuals after subtracting both the \texttt{CANSTAR} activity correction and the Keplerian model (grey). The 10\,\%, 1\,\%, and 0.1\,\% FAP levels are indicated by dashed orange lines. The stellar rotation period (1.96\,d) is marked with a purple dashed line, and the newly derived planetary orbital period (789.92\,d) is marked in light blue (not shown in panel \textbf{(b)} due to its shorter time baseline).}
    \label{fig:correction_tzari}
\end{figure*}

\begin{figure*}
    \centering
    \includegraphics[width=\textwidth]{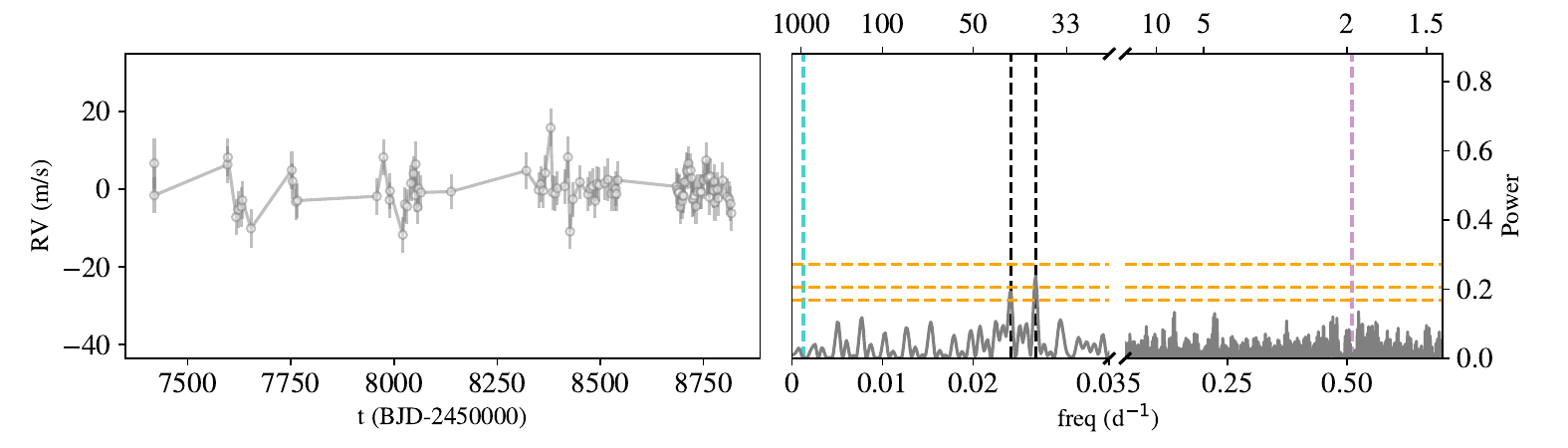}
    \caption{Residual time series (left) and GLS periodogram (right) after subtracting the GP + Keplerian model to the TZ\,Ari RV data (grey). The 10\,\%, 1\,\%, and 0.1\,\% FAP levels are indicated by dashed orange lines. The stellar rotation period (1.96\,d) is shown in purple, the newly derived planetary orbital period (789.92\,d) is in light blue, and the two signals exceeding the 1\,\% FAP level (37.2\,d and 41.4\,d) are marked in black. These likely correspond to stellar activity given the significant (40.18\,d) periodicity observed in the FWHM and $g_2$ indicators.}
    \label{fig:GP_residuals}
\end{figure*}

Figure~\ref{fig:correction_tzari} presents the time series and GLS periodograms of the CARMENES RV observations (blue), together with the \texttt{CANSTAR} activity correction (green), the Keplerian solution (red), and the residuals (grey). In the time series plot (left), the green curve represents the sum of the \texttt{CANSTAR} correction and the Keplerian orbit. This allowed us to visualise the total fit to the observations. The resulting model successfully captures both the Doppler and stellar activity signals present in the data. Notably, the planetary origin of the long-period signal is confirmed by the periodogram (right panel), where the \texttt{CANSTAR} model (green) is shown in isolation and is clearly insensitive to the planetary signal. After applying the correction, the RV residuals show an RRE of 62.4\,\%, measuring the ratio with respect to the variability after subtracting the fitted Keplerian model (Table \ref{tab:statistics_CANSTAR}). The residual GLS periodogram of the \texttt{CANSTAR} correction is largely flat, showing only a minor insignificant peak at 64.7\,d (FAP\,=\,11.7\,\%). In contrast, the residuals of the GP + Keplerian model (Fig.~\ref{fig:GP_residuals}) display two distinct signals above the 1\,\% FAP level at 37.2\,d and 41.4\,d. These periodicities correspond to yearly aliases of one another ($1/37.2 \approx 1/41.4 + 1/365$) and are likely caused by uncorrected stellar activity, given the significant variability observed in the FWHM and $g_2$ indicators at 40.18\,d (Figs.~\ref{fig:TZ_Ari_ClassicalActivity_TimeSeries_LS} and \ref{fig:Reconstruction_timeseries_periodogram_TZ Ari}).

During the final CARMENES observing season, when the star was monitored more intensively, the temporal sampling becomes sufficient to mitigate rotationally modulated stellar activity more effectively with \texttt{CANSTAR}. A zoom-in of this last season is shown in Fig.~\ref{fig:correction_tzari}, where the RRE is further reduced to 33.4\,\%.

\section{Discussion} \label{sec:discussion}

\subsection{The value of orthogonal decomposition}

Standard activity indicators derived from the CCF rely on a simple Gaussian fit. This approach is suboptimal for capturing complex line deformations, requiring ad hoc diagnostics such as the BIS to quantify asymmetries. Furthermore, as shown in Fig.~\ref{fig:CCF_basis_function_fit_EpsEri_TZ_Ari}, the Gaussian model fails to accurately reproduce the non-Gaussian, double-humped CCFs characteristic of many M dwarfs.

A key advantage of the theoretical framework presented here is its generality, allowing for the construction of basis functions tailored to the specific spectral type of the star. We demonstrate that the standard Hermite basis is optimal for describing Gaussian-like CCFs, while M dwarf CCFs exhibiting pronounced double humps are better modelled by the multi-Hermite basis. For the specific case of TZ\,Ari, the humps are largely symmetric, making a two-Gaussian fit sufficient to construct $\boldsymbol{G_0}$. However, more complex or asymmetric CCF morphologies might necessitate three or more Gaussians. A robust and automated approach to determine the optimal number of Gaussians for the mother function is to perform a systematic model selection—for instance, by minimising the Bayesian Information Criterion (BIC)—when fitting the time-averaged template CCF. 

This theoretical framework enables the recovery of the classical CCF activity indicators, as seen in the high level of correlation between $c_1$-CON, $g_2$-FWHM and $g_3$-BIS (Sect. \ref{sec:Aplication_EpsEri} and \ref{sec:Application_TZAri}), in addition to extending the analysis to higher order variability. Moreover, the EVR allowed us to quantify the actual number of distortion basis components needed to describe the observed CCFs.

While a mathematical degeneracy remains between true Doppler shifts induced by planets ($\epsilon$) and apparent shifts caused by stellar activity ($\kappa$), the orthogonal decomposition ensures that our inputs to the NN, the coefficients $g_n(t)$, capture shape information that is mathematically independent of the target variable $l(t)$. Using the first three distortions coefficients, ($c_1$, $g_2$ and $g_3$), we are effectively using similar information to the classical CCF activity indicators, but extending the analysis to higher order coefficients allowed us to improve the activity mitigation performance.

More generally, the distortion coefficients are not limited to the specific architecture proposed. They could serve as inputs for other stellar mitigation techniques, such as the multi-dimensional GP framework. Finally, this parametrisation is computationally efficient and agnostic to the specific pipeline used. It can be easily implemented in standard reduction codes such as \texttt{raccoon} or the ESO DRS.

\subsection{Temporal awareness in activity correction}

The comparison between \texttt{CANSTAR} and the FCN highlights the critical role of temporal information. As shown in Fig.~\ref{fig:Performance_levels_noise}, the FCN, which treats observations as independent snapshots, fails to reach the same correction performance as \texttt{CANSTAR}, particularly in noisy regimes.

Stellar activity is inherently a time-correlated process. Active regions evolve, migrate, and reappear over the stellar rotation period. The inclusion of self-attention mechanisms allows \texttt{CANSTAR} to `learn' these temporal correlations, effectively using the past and future context of the distortion coefficients to mitigate stellar activity occurring at different timescales. This capability is illustrated by the fact that \texttt{CANSTAR} effectively saturates its performance with fewer input coefficients than the FCN. By including the temporal context, the network can extract more information from the lower-order coefficients ($c_1$, $g_2$, $g_3$), reducing the reliance on higher-order terms that are more affected by photon noise.

For exoplanet host stars with sparse RV monitoring, temporal correlations may be lost. In such cases, activity mitigation strategies that do not rely on temporal continuity remain necessary. Our results with the FCN demonstrate the advantage of using the full set of distortion coefficients as inputs compared to classical CCF activity indicators, which are equivalent to using only the lower-order terms ($c_1$, $g_2$, $g_3$).

\subsection{Noise estimation in neural network predictions}

We emphasise the critical importance of correctly propagating photon noise from the spectral level to the CCF and subsequently to the derived activity indicators (see Sect. \ref{sec:Coeff_errors} and Appendix~\ref{sec:appendix_CCF_error}). These precise noise estimates allowed us to inject realistic noise into the synthetic dataset (Sect. \ref{sec:Noise_injection}). Training on data with realistic noise properties is essential to prevent the network from over-fitting to specific noise patterns. Furthermore, these estimates enable us to quantify the uncertainty of the NN predictions. As detailed in Sect. \ref{sec:Results}, we achieve this by evaluating the network on noisy realisations of the input data (propagating measurement error) and by training an ensemble of independent networks to estimate the model variance.

\subsection{Comparison with previous studies}

This study builds upon the previous work developed in \citet{perger2023machine}. We have explored different input data, transitioning from classical CCF activity indicators (CON, FWHM and BIS) to time series of distortion coefficients, and different NN architectures, from a CNN to a convolutional attention network. Both studies relied on \texttt{StarSim} data for training. However, in this work, we utilised a newer version of the code that has been improved, both in its physical model as well as on the practical implementation (Gomes et al., in prep.). 

\begin{figure*}
    \centering
    \includegraphics[width=\linewidth]{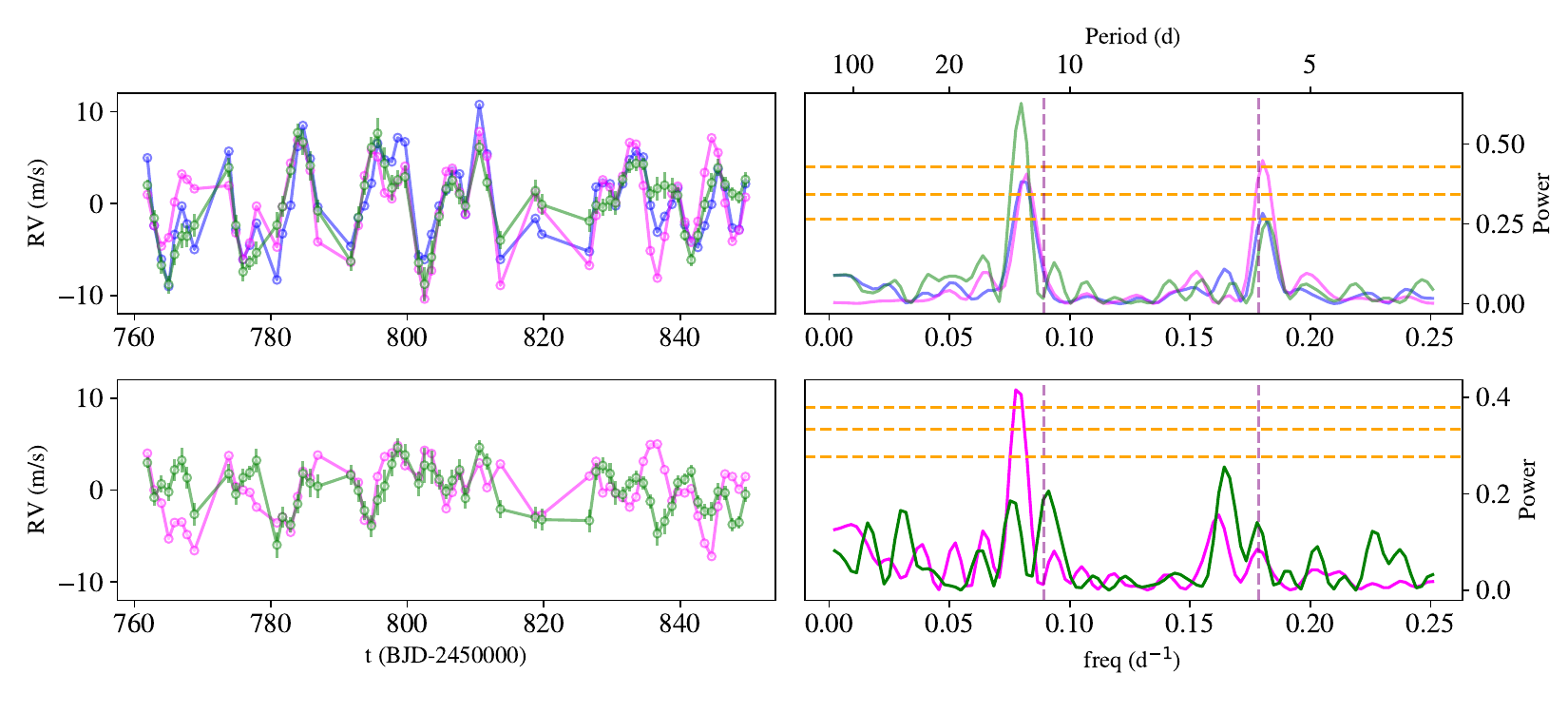}
    \caption{Top: Radial velocity time series (left) and GLS periodogram (right) of the HARPS $\epsilon$\,Eri data (blue) compared to the stellar activity signal predicted by \texttt{CANSTAR} (green) and \citet[ magenta]{perger2023machine}. We note that the \citet{perger2023machine} did not provide uncertainties to their NN predictions. Bottom: Residuals after subtracting the prediction with their corresponding periodogram (right). The 10\,\%, 1\,\%, and 0.1\,\% FAP levels are indicated by dashed orange lines.}
    \label{fig:correction_comparison_perger}
\end{figure*}

We can directly compare the performance of \citet{perger2023machine} against \texttt{CANSTAR} regarding stellar activity correction on the $\epsilon$\,Eri dataset (Fig.~\ref{fig:correction_comparison_perger}). The \texttt{CANSTAR} correction matches the observed data better, resulting in an RRE reduction to 52.5\,\% of the original variability, compared to the 67\,\% reduction level of the \citet{perger2023machine} results\footnote{We note that \citet{perger2023machine} reported a reduction value of 45\,\%. However, this corresponds to the reduction in MSE. The corresponding reduction in RRE is $\sqrt{0.45} \approx 67$\,\%.}. The residuals from both approaches show significant differences. The GLS periodogram of the \texttt{CANSTAR} residuals shows a clear mitigation of the rotational period (FAP\,>\,10\,\%). On the other hand, the peak at the rotation period remains highly significant (FAP\,<\,0.1\,\%) in the GLS periodogram of the \citet{perger2023machine} residuals. The residuals from both methods display a low degree of correlation, with a Pearson correlation coefficient of 0.38.

In Sect.~\ref{sec:Results TZ Ari}, we compared the \texttt{CANSTAR} activity correction on TZ\,Ari against the GP framework with the SHO kernel from \citet{quirrenbach2022carmenes}, but we limited the analysis to CARMENES. \texttt{CANSTAR} achieves compatible but more precise determinations of the semi-amplitude and period of the Keplerian parameters (Fig.~\ref{fig:comparison_GP}). The GP achieves a larger reduction of RRE, as illustrated in the lower fitted jitter term. This is expected due to the flexibility of GP modelling solely on RV data, which can lead to over-fitting leaving small residual amplitudes \citep{blunt2023overfitting}. However, two residual signals in the GLS periodogram remain above the 1\,\% FAP level after the GP solution, with periods 37.2\,d and 41.4\,d. These periods are likely yearly aliases of a stellar activity signal, as evidenced by significant variability in the $g_2$ and FWHM indicators at 40.18\,d (Table \ref{tab:Appendix_TZ_Ari}). In contrast, \texttt{CANSTAR} effectively removes these signals because its correction is explicitly conditioned on the line-shape distortion coefficients. This demonstrates that \texttt{CANSTAR} provides a physically motivated correction that avoids the over-fitting pitfalls of pure RV modelling.

\subsection{Bridging the synthetic gap} \label{sec:Synthetic gap}

We find a difference in performance when evaluating the \texttt{CANSTAR} network on synthetic versus real data. In our noiseless simulations, \texttt{CANSTAR} achieves nearly perfect correction, and even in datasets with injected noise matching the observations, the network achieves an RRE of 12.9\,\% for $\epsilon$\,Eri and 35.7\,\% for TZ\,Ari. However, when applied to the real datasets, the RRE becomes 52.5\,\% and 61.6\,\%, respectively. This performance gap indicates that while \texttt{StarSim} provides a physically realistic output, it still lacks certain stellar activity and instrumental effects present in reality. 

First, our noise injection procedure currently accounts only for photon noise measured at the spectral level (white noise). However, real observations are affected by instrumental systematics and atmospheric effects that introduce correlated noise into the data. Since our network is trained solely on white noise, it may be less effective at filtering out these complex, non-Gaussian systematic trends. Future synthetic datasets should therefore integrate instrumental systematics to better simulate the red noise floor of the specific spectrograph.

Second, current \texttt{StarSim} simulations rely on 1D PHOENIX stellar spectra \citep{husser2013new} combined with modified bisector line shapes to simulate the photosphere, spots, and faculae. However, spots and faculae are direct manifestations of the interplay between magnetic fields and the complex 3D structure of the photosphere \citep{witzke2022can}. This dependence on 1D atmospheric models is a shared limitation among current simulators, such as SOAP-GPU \citep{zhao2023soap} and SOAP 4.0 \citep{cristo2025soapv4}.

A significant step towards bridging this physical gap is the use of synthetic spectra derived from 3D magnetohydrodynamic (MHD) simulations. The upcoming version of \texttt{StarSim} (Gomes et al. in prep., Stucki et al. in prep.) is designed to incorporate spectra from the MURaM MPS-ATLAS 3D MHD models \citep{witzke2024testing}. By explicitly modelling the 3D structure of the stellar atmosphere driven by convection, this approach will improve the physical prescription of the photosphere, spots and faculae, while enabling the inclusion of other convective phenomena such as granulation and super-granulation. Simulating these phenomena will be essential for extending this framework to less active stars \citep{reinhold2019transition,meunier2020effects}.

On the other hand, accurately modelling very active, flaring M dwarfs would require \texttt{StarSim} to simulate coronal heating driven by high-intensity magnetic fields. The self-consistent simulation of stellar chromospheres and flare events remains a significant limitation shared across the entire field \citep{allen2026jwst}.

\section{Conclusions} \label{sec:Conclusions}

We have introduced \texttt{CANSTAR}, a novel framework that exploits information on CCF line-shape distortions and their temporal evolution. The framework consists of two main steps: (1) separating pure Doppler shifts from line-shape distortions using an orthonormal basis expansion, which enables the subtraction of distortion coefficients time series, and (2) modelling the temporal evolution of these stellar activity distortions with a convolutional attention network trained on synthetic datasets generated with \texttt{StarSim}.

We have demonstrated that \texttt{CANSTAR} achieves near-perfect stellar activity correction on simulated data, outperforming an FCN that does not explicitly model temporal correlations. The performance worsens when noise resembling that of real observations is injected, primarily because higher-order coefficients become increasingly noise dominated.

When trained on synthetic noisy data and applied to real observations of $\epsilon$\,Eri and TZ\,Ari, \texttt{CANSTAR} successfully mitigated a significant fraction of the stellar activity signal. For $\epsilon$\,Eri, we improved the activity correction provided in \citet{perger2023machine} by reducing the RRE from 67\,\% to 52.5\,\%. Applying the \texttt{CANSTAR} correction improves the detection limit for a 10\,\% relative precision in $K$ to $2.51 \pm 0.43$\,m\,s$^{-1}$, compared to the baseline of $3.99 \pm 0.18$\,m\,s$^{-1}$ obtained for the uncorrected (null hypothesis) case. 

For TZ\,Ari, the network successfully disentangles planetary Doppler shifts from activity-induced signals, yielding improved orbital parameter estimates for TZ\,Ari\,b compared to a GP fit. This demonstrates that \texttt{CANSTAR} outperforms the current state-of-the-art solution, enabling both robust activity correction and precise planetary characterisation. After subtracting the Keplerian solution, \texttt{CANSTAR} achieves an RRE of 62.4\,\% on the full dataset. Notably, this residual scatter is further reduced to 33.4\,\% for the final observing season, where the higher cadence allowed for proper sampling of the 1.96\,d stellar rotation period. 

Beyond the integrated framework, the individual components of \texttt{CANSTAR} offer independent utility. The distortion coefficients obtained from the orthonormal basis expansion can reproduce classical activity indicators while providing sensitivity to additional stellar phenomena via higher-order terms. Furthermore, the convolutional attention network has a general time–aware architecture capable of capturing variability on multiple timescales. Its inputs are not limited to CCF coefficients, but it can be readily adapted to use photometry, chromospheric indices, or line-by-line activity indicators.

Looking forward, there is a clear path to bridge the performance gap between simulations and real data. Future \texttt{StarSim} developments will incorporate spectra from 3D MHD simulations, allowing us to model less active stars where granulation and faculae dominate. By integrating these physical effects along with instrumental systematics into the training sets, we expect to further enhance the network's predictive power.

Ultimately, this work demonstrates that NNs are not just a future prospect but a present capability, and they already outperform established state-of-the-art solutions, such as GPs, in regimes dominated by complex stellar activity. By effectively using temporal context and high-order line shape distortions, \texttt{CANSTAR} offers a robust pathway to disentangling planetary signals from stellar activity effects. This approach, with further maturation and work closing the gap between synthetic and real data, can be instrumental in pushing detection limits towards the photon-noise floor, which would enable the discovery of lower-mass planets and eventually unlock the domain of exo-Earths.

\begin{acknowledgements}

The authors thank the anonymous referee and the editor for their constructive feedback and careful reading of the manuscript. We also warmly thank Pedro Figueira for providing valuable comments that helped improve this work. J.B.-P., M.P., G.A.-E., I.R., J.C.M., O.P. and S.S. acknowledge financial support from Spanish grants PID2021-125627OB-C31 funded by MCIU/AEI/10.13039/501100011033 and by “ERDF A way of making Europe”, PID2024-158486OB-C31 funded by MCIU/AEI, by the programme Unidad de Excelencia María de Maeztu CEX2020-001058-M and by the MaX-CSIC Excellence Award MaX4-SOMMA-ICE, by the Generalitat de Catalunya/CERCA programme, and by the European Research Council (ERC) under the European Union’s Horizon Europe  programme (ERC Advanced Grant SPOTLESS; no. 101140786). Views and opinions expressed are however those of the author(s) only and do not necessarily reflect those of the European Union or the European Research Council. Neither the European Union nor the granting authority can be held responsible for them. J.B.-P., M.P. and G.A.-E. also acknowledge financial support from Spanish grant PID2020-120375GB-I00, funded by MCIU/AEI, and Consolidación 2022 CNS2022-136050. J.B.-P. also acknowledges financial support from Spanish grant PRE2022-101942 funded by MICIU/AEI/10.13039/501100011033 and ESF+. M.L acknowledges support by the UKRI (Grant EP/X027562/1). The data production, processing and analysis tools for this paper have been developed, implemented and operated in collaboration with the Port d’Informaci\'o Cient\'ifica (PIC) data center. PIC is maintained through a collaboration agreement between the Institut de F\'isica d’Altes Energies (IFAE) and the Centro de Investigaciones Energ\'eticas, Medioambientales y Tecnol\'ogicas (CIEMAT).

\end{acknowledgements}

\bibliographystyle{aa}
\bibliography{bibliography} 

\clearpage
\newpage

\begin{appendix}

\section{TESS photometry} \label{sec:TESS}

The Transiting Exoplanet Survey Satellite (TESS) is performing an all-sky survey in search for transiting exoplanets around the closest and brightest stars \citep{ricker2015transiting}. TESS first observed TZ\,Ari in Sectors 70 and 71 at the start of the mission's year 6, spanning from September to November 2023. We analyse the Pre-search Data Conditioning Single Aperture Photometry (PDCSAP) flux provided by the mission, which has been corrected for trends specific to each CCD and observing sector. The light curves and GLS periodograms are dominated by a stable signal induced by the rotation period. We also see a large flare event at around BJD\,$\approx$\,2460210.

Based on the identification of the highest power peak in the periodograms of the two sectors and estimating the uncertainty from the FWHM of the dominant peak, we report a rotation period of 1.96$\pm$0.07\,d, which is in agreement with the reported period from high-resolution spectroscopy and ground-based photometry analysis in \citet{quirrenbach2022carmenes}.

\begin{figure*} 
    \centering
    \includegraphics[width=\textwidth]{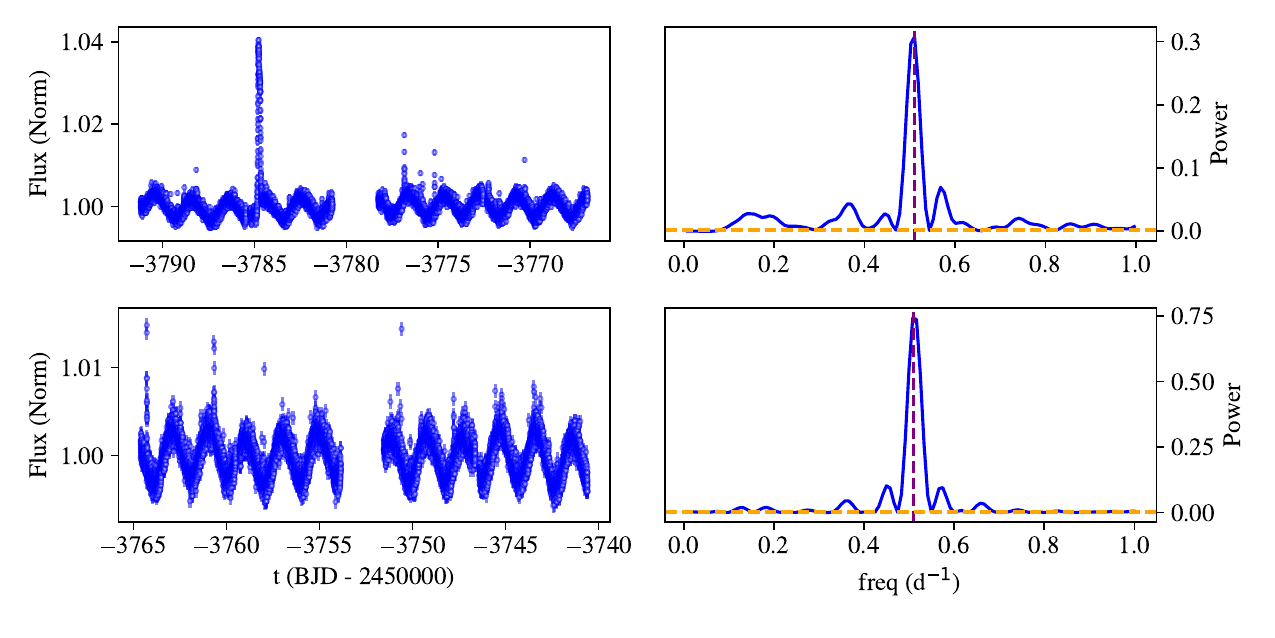}
    \caption{Time series (left) and GLS periodograms (right) of the TESS photometry of TZ\,Ari for sectors 70 (top) and 71 (bottom). We mark in purple the stellar rotation period (1.96$\pm$0.07\,d).}
    \label{fig:TESS}
\end{figure*}

\section{Statistics of CCF products} \label{sec:appendix_table}

We show the main statistics on the derived distortion coefficients and classical CCF activity indicators of the test targets of the study $\epsilon$\,Eri and TZ\,Ari. These statistics consist of the root mean square (rms) of the time series, the mean error, the EVR, the cumulative EVR, and the statistics from the pre-whitening process, consisting of the RRE, the identified periods and the fitted amplitudes.

\begin{table*}[]
    \centering
    \caption{Statistics on the coefficients derived of $\epsilon$\,Eri and the corresponding ones for the classical CCF indicators.}
    \begin{tabular}{ccccccccc}
    \hline \hline
    & & & & & Cumulative & \multicolumn{3}{c}{Pre-whitening statistics} \\ \cline{7-9}
    indicator & units & RMS & error & EVR (\%) & EVR (\%) & RRE (\%) & Periods (d) & Amplitudes \\
    \hline \hline
$c_0$ & $\times 10^{3}$ & 0.476  & 0.085 & 5.0    & 5.0 & 52.6 & 11.19, 5.54 & 0.53, 0.18 \\
$c_1$ & $\times 10^{3}$ & 0.832  & 0.073 & 35.4    & 40.4 & 38.6 & 12.46, 10.65, 5.54 & 0.94, 0.45, 0.29 \\
$l$ & $\times 10^{2}$ & 0.112  & 0.008 & 18.3    & 58.7 & 60.1 & 12.12, 5.54 & 0.10, 0.08 \\
$g_2$ & $\times 10^{2}$ & 0.116  & 0.008 & 14.5    & 73.1 & 29.0 & 11.79, 5.54, 137.36, 12.83 & 0.13, 0.06, 0.06, 0.04 \\
$g_3$ & $\times 10^{3}$ & 0.705  & 0.081 & 9.5    & 82.7 & 37.5 & 12.12, 5.54, 10.65 & 0.69, 0.54, 0.29 \\
$g_4$ & $\times 10^{3}$ & 0.215  & 0.082 & 6.1    & 88.7 & 78.8 & 12.83 & 0.19 \\
$g_5$ & $\times 10^{3}$ & 0.165  & 0.083 & 4.1    & 92.9 & 71.8 & 12.12 & 0.17 \\
$g_6$ & $\times 10^{3}$ & 0.121  & 0.084 & 2.3    & 95.2 & 74.0 & 11.79, 137.36 & 0.08, 0.08 \\
$g_7$ & $\times 10^{3}$ & 0.122  & 0.084 & 1.7    & 96.9 & 85.9 & 11.79 & 0.09 \\
$g_8$ & $\times 10^{3}$ & 0.100  & 0.084 & 0.9    & 97.8 & 100 & -- & -- \\
RV & m\,s$^{-1}$ & 4.691 & 0.286 & -- & --     & 58.5 & 12.12, 5.54 & 4.14, 3.51 \\
CON & \% & 0.058 & 0.004 & -- & --     & 58.3 & 12.12 & 0.07 \\
FWHM & m\,s$^{-1}$ & 13.950 & 0.872 & -- & --     & 74.5 & 11.79 & 13.04 \\
BIS & m\,s$^{-1}$ & 5.472 & 0.294 & -- & --     & 34.6 & 12.12, 5.54, 10.65, 3.56 & 5.55, 4.03, 2.01, 1.32 \\
\hline
    \end{tabular}
    \label{tab:Appendix_EpsEri}
\end{table*}

\begin{table*}[]
    \centering
    \caption{Statistics on the coefficients derived of TZ\,Ari and the corresponding ones for the classical CCF indicators.}
    \begin{tabular}{ccccccccc}
    \hline \hline
    & & & & & Cumulative & \multicolumn{3}{c}{Pre-whitening statistics} \\ \cline{7-9}
    indicator & units & RMS & error & EVR (\%) & EVR (\%) & RRE (\%) & Periods (d) & Amplitudes \\
    \hline \hline
$c_0$ & $\times 10^{2}$ & 0.462  & 0.061 & 5.0    & 5.0 & 94.5 & 18.30 & 0.22 \\
$c_1$ & $\times 10^{2}$ & 0.262  & 0.063 & 35.4    & 40.4 & 100.0 & -- & -- \\
$l$ & $\times 10^{2}$ & 0.445  & 0.060 & 18.3    & 58.7 & 60.9 & 1.96, 194.44, 1.95 & 0.38, 0.26, 0.17 \\
$g_2$ & $\times 10^{2}$ & 0.164  & 0.063 & 14.5    & 73.1 & 74.8 & 2.05, 40.18, 1.26 & 0.11, 0.08, 0.07 \\
$g_3$ & $\times 10^{2}$ & 0.132  & 0.064 & 9.5    & 82.7 & 72.0 & 1.96, 1.95 & 0.11, 0.06 \\
$g_4$ & $\times 10^{3}$ & 0.781  & 0.643 & 6.1    & 88.7 & 100.0 & -- & -- \\
$g_5$ & $\times 10^{3}$ & 0.710  & 0.642 & 4.1    & 92.9 & 100.0 & -- & -- \\
$g_6$ & $\times 10^{3}$ & 0.905  & 0.642 & 2.3    & 95.2 & 86.1 & 71.36, 184.19 & 0.55, 0.39 \\
$g_7$ & $\times 10^{3}$ & 0.649  & 0.642 & 1.7    & 96.9 & 100.0 & -- & -- \\
$g_8$ & $\times 10^{3}$ & 0.664  & 0.646 & 0.9    & 97.8 & 100.0 & -- & -- \\
RV & m\,s$^{-1}$ & 15.215 & 2.192 & -- & --     & 61.2 & 1.96, 194.44, 1.95 & 0.01, 0.01, 0.01 \\
CON& \% & 0.086 & 0.029 & -- & --     & 95.9 & 1.97 & 0.03 \\
FWHM & m\,s$^{-1}$ & 13.978 & 10.887 & -- & --     & 87.5 & 40.18 & 0.01 \\
BIS & m\,s$^{-1}$ & 7.374 & 5.734 & -- & --     & 72.0 & 1.96, 1.95 & 0.01, 0.00 \\
\hline
    \end{tabular}
    \label{tab:Appendix_TZ_Ari}
\end{table*}

\section{CCF error propagation} \label{sec:appendix_CCF_error}

The noise budget of the CCF is determined by the uncertainty of the measurement of electrons on the spectrograph detector. The CCF value at a given RV point is computed as the weighted sum of the flux from $N_p$ pixels that overlap with the $N_m$ spectral lines defined in the weighted mask. This can be expressed as
\begin{equation} \label{eq:CCF}
    \mathrm{CCF}(v) = \sum_{l=1}^{N_m}\sum_{x=1}^{N_p} m_l f_x \Delta_{xl},
\end{equation}
where $m_l$ is the weight of the mask for spectral line $l$, $f_x$ represents the flux of pixel $x$ in the analog-to-digital Unit (ADU), and $\Delta_{xl}$ is the fraction of pixel $x$ covered by the mask line $l$  after shifting it by a $v$ shift.

To estimate the uncertainty on the CCF profile, we consider the noise variance of each individual pixel. The flux in ADU, $f_x$, is related to the number of photo-electrons, $n_{e^-}$, via the detector gain $g$ (in $\mathrm{ADU}/e^-$):
\begin{equation} \label{eq:fx}
    f_x = g \cdot n_{e^-}.
\end{equation}

Given that the photo-electrons follow Poisson statistics, the uncertainty in the electron count is $\delta n_{e^-} = \sqrt{n_{e^-}}$. Converting this to ADU and including the detector readout noise, $\mathrm{RON}$, the total variance for a single pixel $x$ is given by:
\begin{equation} \label{eq:fx^2}
    \delta f_x^2 = g^2 \delta n_{e^-}^2 + g^2 \mathrm{RON}^2 = g^2 n_{e^-} + g^2 \mathrm{RON}^2 = g f_x + g^2 \mathrm{RON}^2.
\end{equation}

Since the noise contributions from individual pixels are statistically independent, the total error on the CCF is obtained by adding the individual variances in quadrature. Applying standard error propagation to Eq.~(\ref{eq:CCF}) results in
\begin{equation} \label{eq:deltaCCF_1}
    \delta \mathrm{CCF} = \sqrt{ \sum_{l=1}^{N_m}\sum_{x=1}^{N_p} \left( m_l \Delta_{xl} \right)^2 \delta f_x^2 }.
\end{equation}
Substituting the expression for pixel variance:
\begin{equation} \label{eq:deltaCCF_2}
    \delta \mathrm{CCF} = \sqrt{ \sum_{l=1}^{N_m}\sum_{x=1}^{N_p} \left( m_l \Delta_{xl} \right)^2 \left( g f_x + g^2 \mathrm{RON}^2 \right) }
\end{equation}

\section{Correlation between distortion coefficients and classical CCF activity indicators}

We show the correlation between decomposition coefficients and classical CCF indicators.

For the $\epsilon$\,Eri case, the strongest correlations are CON with $c_1$, RV with $l$, FWHM with $g_2$, and BIS with $g_3$. This confirms that the decomposition recovers the classical CCF activity indicators. The strongest correlation between decomposition coefficients is between $l$ and $g_3$, and between $c_1$ and $g_2$.

Similarly, for the TZ\,Ari case, the strongest correlations are also between CON with $c_1$, RV with $l$, FWHM with $g_2$, and BIS with $g_3$. However, we see a smaller degree of correlation between FWHM and $g_2$, and between BIS and $g_3$, compared to the $\epsilon$\,Eri case. The strongest correlation between decomposition coefficients is between $c_0$ and $c_1$, and between $l$ and $g_3$.

\begin{figure*}
    \centering
    \includegraphics[width=0.8\textwidth]{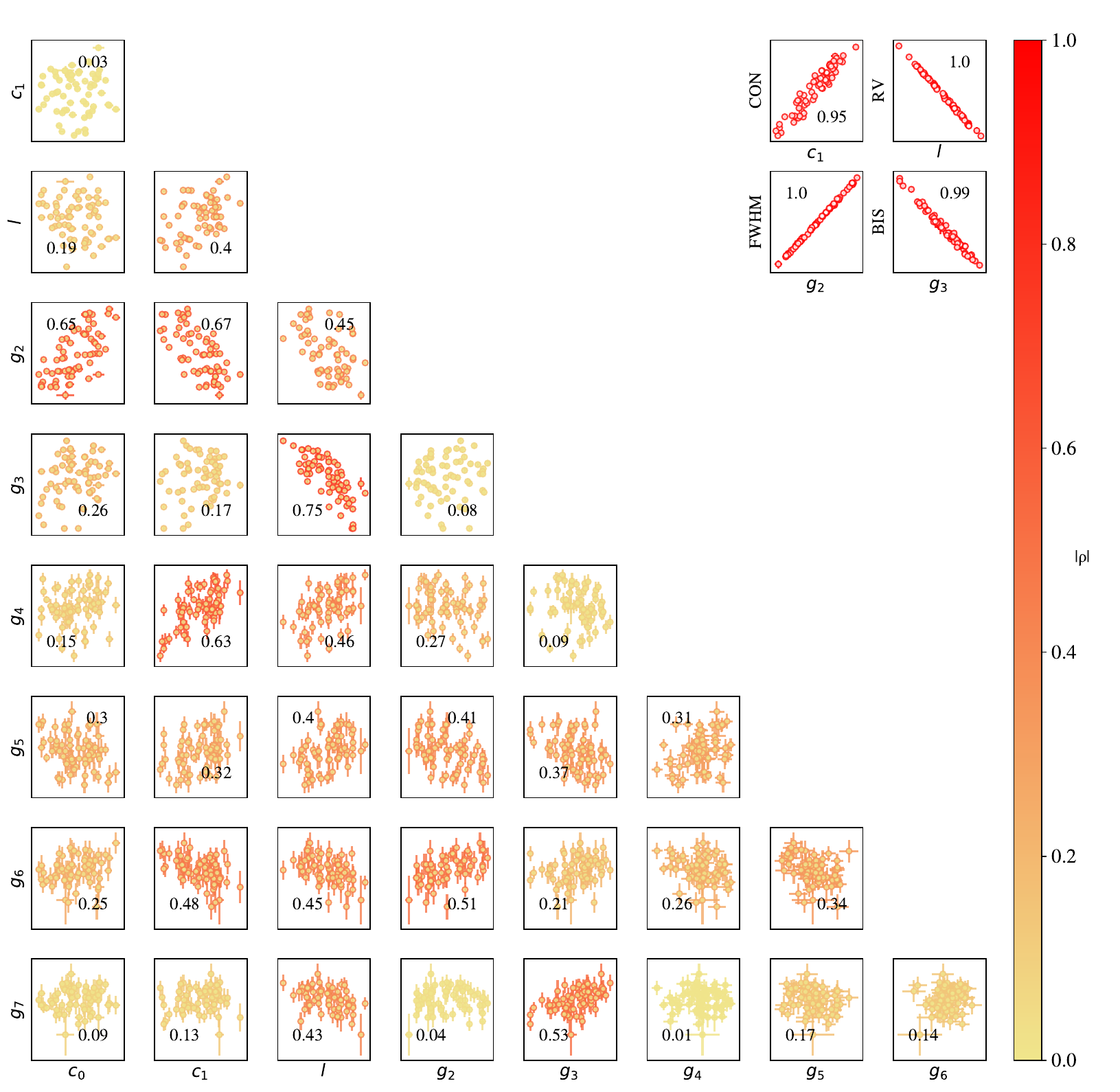}
    \caption{Correlations between the first 8 decomposition coefficients, colour-coded according to the absolute value of their Pearson correlation coefficient, where the exact value is printed over the image. Strongest correlations with classical CCF indicators appear at CON–$c_1$, RV–$l$, FWHM–$g_2$, and BIS–$g_3$ (top right).}
    \label{fig:correlation_plot_EpsEri}
\end{figure*}

\begin{figure*}
    \centering
    \includegraphics[width=0.8\textwidth]{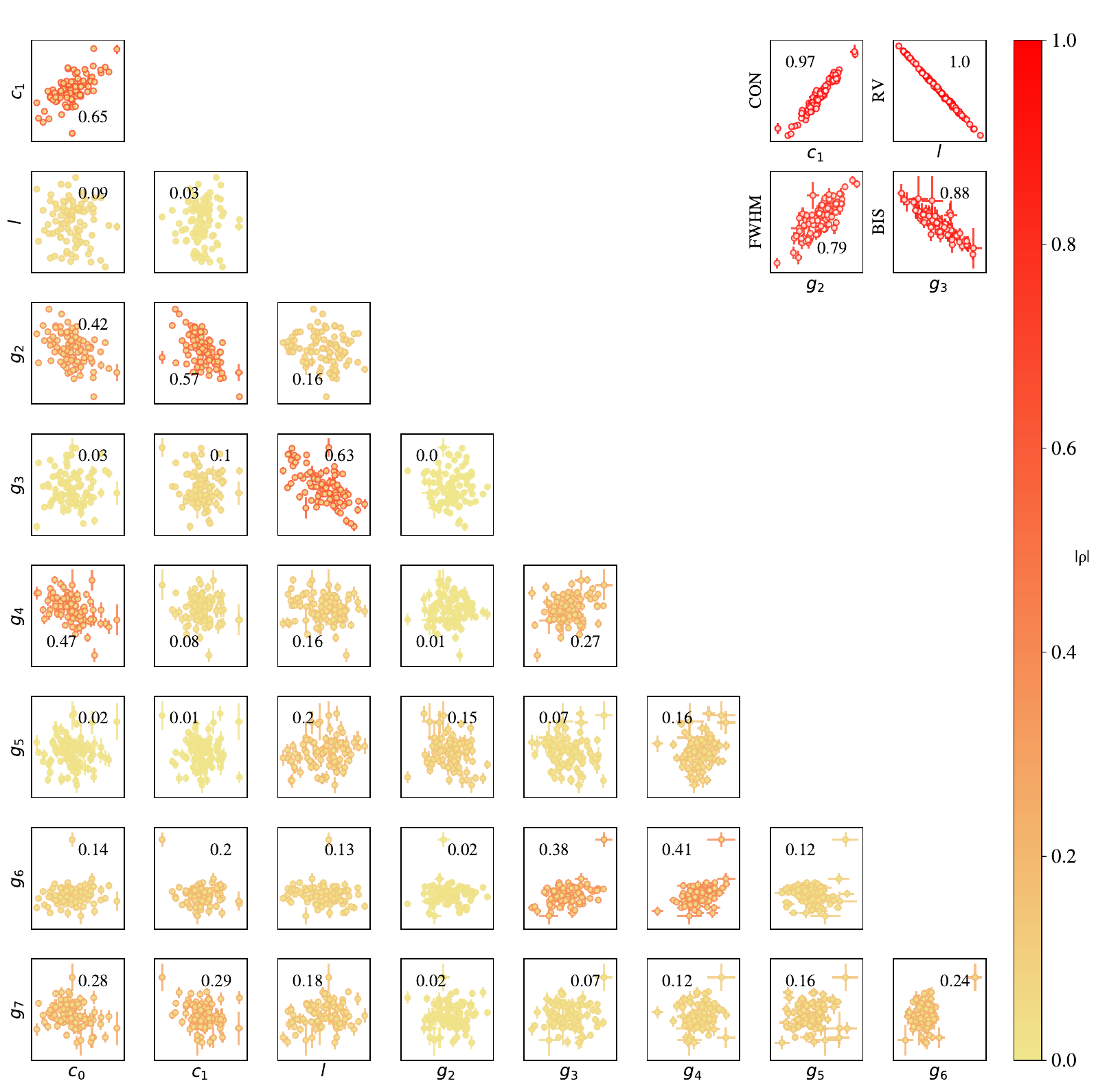}
    \caption{Same as Fig.~\ref{fig:correlation_plot_EpsEri} but for the multi-Hermite basis applied to TZ\,Ari.}
    \label{fig:correlation_plot_TZ Ari}
\end{figure*}

\section{Scale mismatch between synthetic and observed RMS time series coefficients} \label{sec:appendix_scale_coeffs}

We generate synthetic datasets using \texttt{StarSim} aimed at reproducing the stellar activity signals present in the observed data. However, a discrepancy remains between the simulations and observations, as illustrated by the difference in RMS between the synthetic and observed coefficient time series. In order to minimise this discrepancy, we apply $z$-score normalisation to both datasets so that their scales matches.

\begin{figure*}
    \centering
    \includegraphics[width=\textwidth]{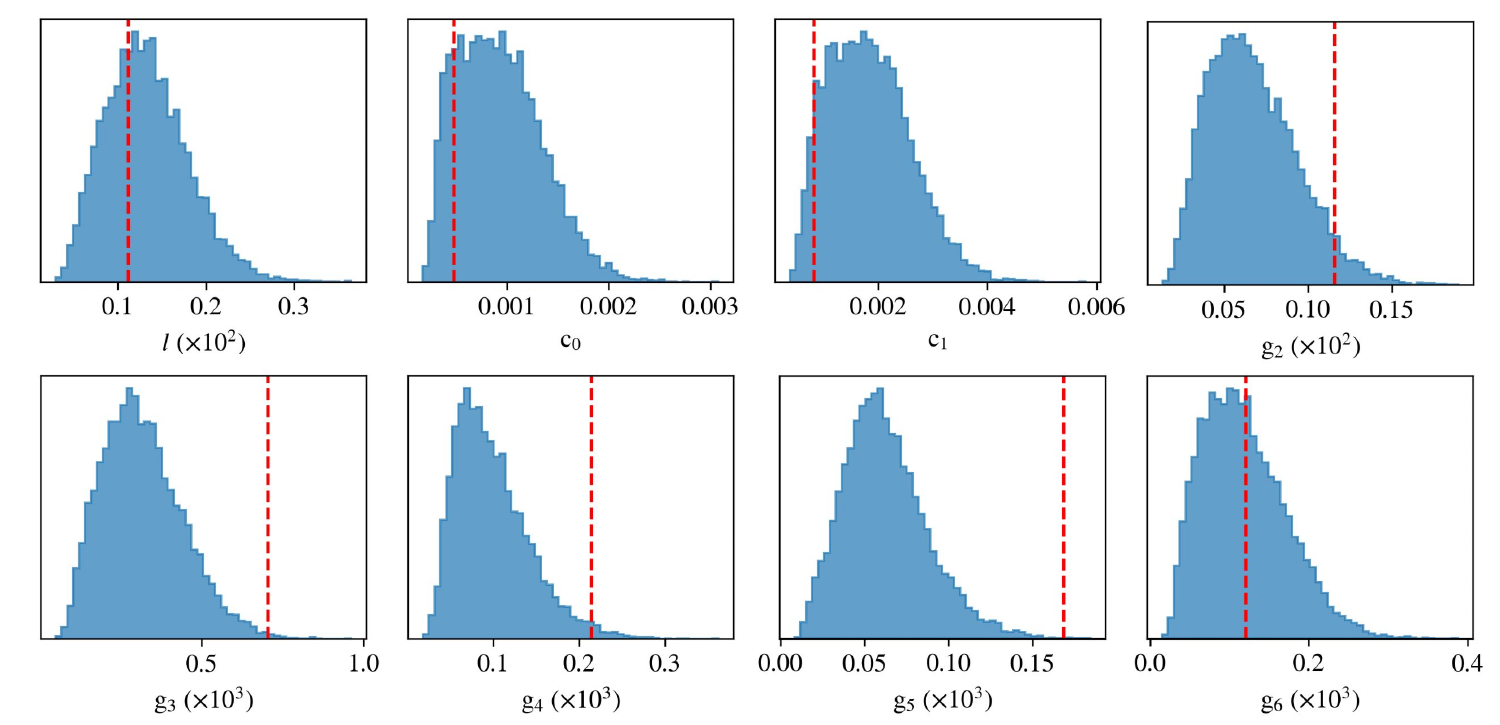}
    \caption{Histograms showing the distribution of RMS values for the coefficients derived from 10,000 \texttt{StarSim} simulations for $\epsilon$\,Eri. The RMS values of the actual observed HARPS coefficients are shown by vertical red dashed lines. We apply z-score normalisation to bridge the gap between the time series RMS from the simulated distributions and the observed ones.}
    \label{fig:coeff_mismatch}
\end{figure*}

\section{Injection and retrievals for the null hypothesis} \label{sec:injection_retrieval_appendix}

We perform injection and retrievals for the null hypothesis case. The detection limits are $3.99 \pm 0.18$\,m\,s$^{-1}$ (10\,\% precision in $K$) and $3.87 \pm 0.18$\,m\,s$^{-1}$ (30\,\% precision).

\begin{figure*} 
    \centering
    \includegraphics[width=\textwidth]{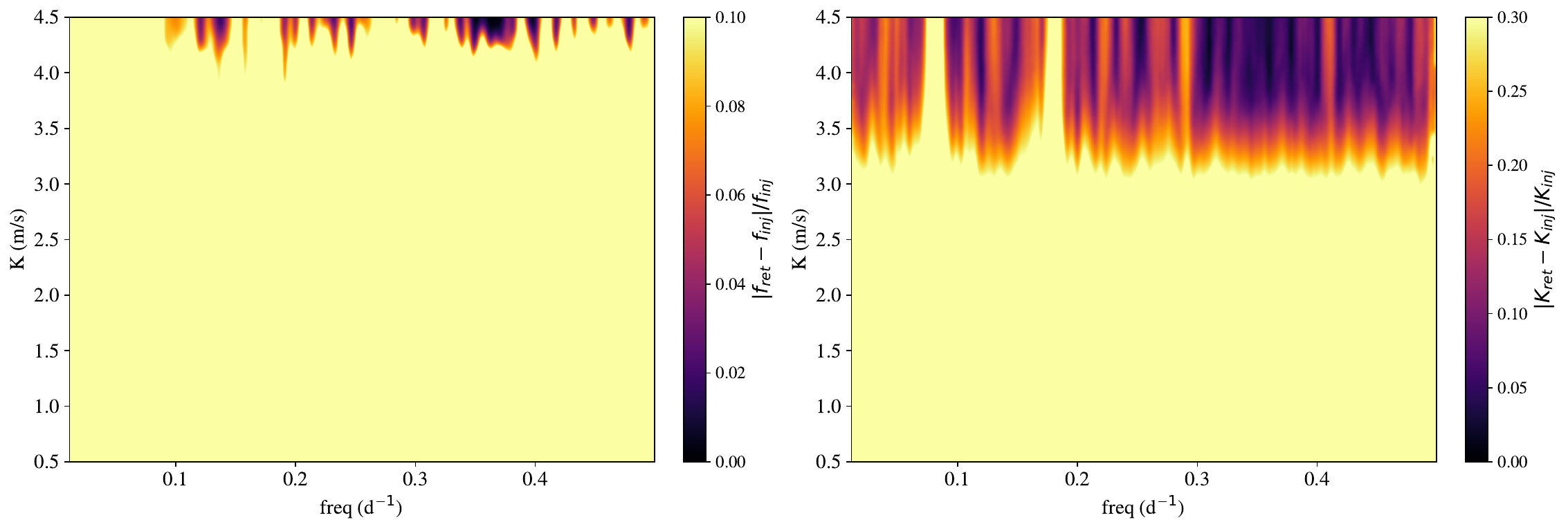}
    \caption{Same as Fig.~\ref{fig:detection_limits} but for the null hypothesis case.}
    \label{fig:injection_retrieval_appendix}
\end{figure*}

\section{Gaussian process kernel} \label{sec:appendix_GP}

In order to benchmark the stellar activity correction capabilities of \texttt{CANSTAR}, we also test the GP regression strategy used in \citet{quirrenbach2022carmenes}. In agreement with such work, we employ the Simple Harmonic Oscillator (SHO) kernel. This kernel is physically motivated, describing a stochastically driven, damped harmonic oscillator, which acts as an effective approximation for the quasi-periodic variability characteristic of stellar rotation and active region evolution. This kernel takes the form
\begin{equation}
k(\tau) = S\omega Q \exp\left(-\frac{\omega\tau}{2Q}\right) \left[ \cos(\eta \omega \tau) + \frac{1}{2Q\eta} \sin(\eta \omega \tau) \right],
\label{eq:SHO_kernel}
\end{equation}
where $S$ is the amplitude, $\omega$ is the angular frequency, and $Q$ is the quality factor of the oscillator. $\omega$ is related to the rotational period $\omega=2\pi/P$ and $Q$ is also related to the typical coherent lifetime, $l$ of the signal, $Q=l\omega/2$. In our analysis, we fit for the hyper-parameters utilising the \texttt{celerite2} implementation \citep{foreman2017fast}.

\section{{\tt CANSTAR} + Keplerian solution posteriors} \label{sec:appendix_posteriors}

We show the posterior distributions from the {\tt CANSTAR} + Keplerian model.

\begin{figure*} 
    \centering
    \includegraphics[width=\textwidth]{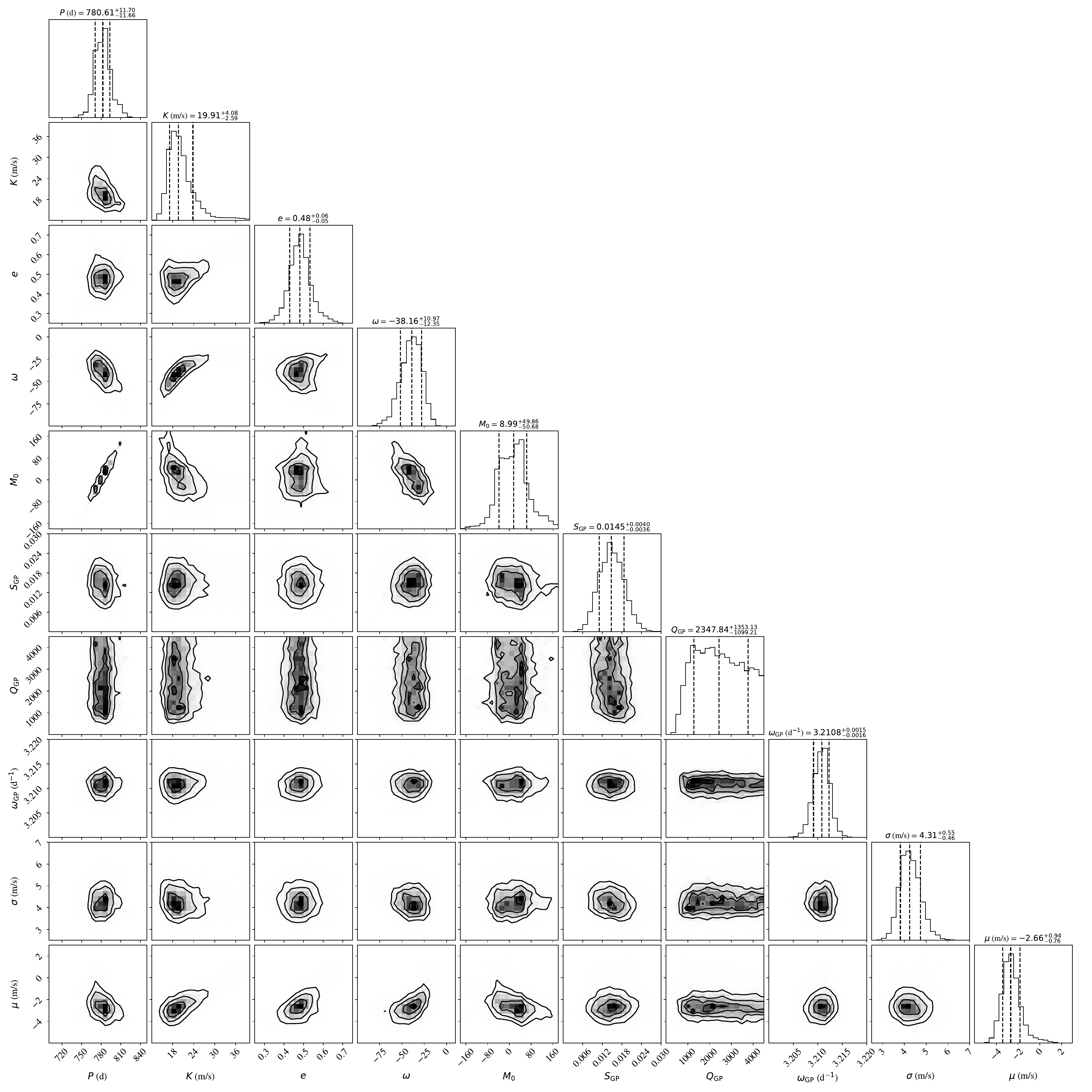}
    \caption{Posterior distribution from \texttt{emcee} for the {\tt CANSTAR} + Keplerian solution}
    \label{fig:corner_plot_canstar}
\end{figure*}

\end{appendix}

\end{document}